\documentclass{amsart}

\usepackage{amsfonts,amssymb,amsmath,latexsym}
\usepackage{graphicx}

\newcommand{\onlinecite}{\cite}

\numberwithin{equation}{section}

\newtheorem{theo}{Theorem}[section]
\newtheorem{defini}[theo]{Definition}
\newtheorem{proposi}[theo]{Proposition}
\newtheorem{lemma}[theo]{Lemma}

\newcommand{\CC}{{\mathbb C}}

\newcommand{\RR}{{\mathbb R}}

\newcommand{\ZZ}{{\mathbb Z}}

\newcommand{\Gg}{\mathcal{G}}
\newcommand{\Ww}{\mathcal{W}}
\newcommand{\Ss}{\mathcal{S}}

\newcommand{\Tt}{\mathcal{T}}

\newcommand{\Cc}{\mathcal{C}}
\newcommand{\Jj}{\mathcal{J}}

\newcommand{\Uu}{\mathcal{U}}
\newcommand{\Vv}{\mathcal{V}}

\newcommand{\one}{{\bf 1}}
\newcommand{\nul}{{\bf 0}}

\newcommand{\comm}[1]{}
\newcommand{\comment}[1]{}

\newcommand{\diag}{{\rm diag}}

\newcommand{\bgM}{\left(\begin{matrix}}
\newcommand{\enM}{\end{matrix}\right)}

\newcommand{\lgth}{ L} 
\newcommand{\wdth}{ N} 

\begin{document}

\title[Relations between Transfer and Scattering Matrices]
{Relations between Transfer and Scattering Matrices in the presence of Hyperbolic Channels}

\author{Christian Sadel}
\email{csadel@math.uci.edu}
\address{University of California, Irvine,
Department of Mathematics,
Irvine, CA 92697-3875,  USA}


\begin{abstract}
We consider a cable described by  a discrete, 
space-homogeneous, quasi one-dimensional Schr\"odinger  operator $H_0$.
We study the scattering by a finite disordered piece (the scatterer) inserted inside this cable.
For  energies $E$ where  $H_0$ has only elliptic channels 
we use
 the Lippmann-Schwinger equations to show that the scattering matrix and
the transfer matrix, written in an appropriate basis, are related by a certain polar decomposition.  For energies $E$ where
 $H_0$ has hyperbolic channels  we show that the  scattering matrix is related to
a reduced transfer matrix and both are of smaller dimension than the transfer matrix. 
Moreover, in this case the scattering matrix is determined from a limit of larger dimensional scattering matrices, as follows:
We take a piece of the cable of length $m$, followed by the scatterer and another piece of the cable
 of length $m$, consider the scattering matrix of these three joined pieces inserted inside an ideal lead at energy $E$
(ideal means only elliptic channels),
and take the limit $m\to\infty$.
\end{abstract}

\maketitle

\section{Introduction}
\label{sec-int}

We consider discrete quasi one-dimensional Schr\"odinger operators on strips of width $\wdth$ of the form
\begin{equation} \label{eq-H-int}
(H \Psi)_n
\,=\,
-\Psi_{n+1}\,-\,\Psi_{n-1}\,+\,V_n \Psi_n\;
\end{equation}
where $\Psi=(\Psi_n)_{n\in\ZZ} \,\in\,\ell^2(\ZZ,\CC^\wdth)\cong\ell^2(\ZZ)\otimes\CC^\wdth$ 
is an $\ell^2$ sequence of vectors in $\CC^\wdth$ and 
$V_n\in{\rm Her}(\wdth)$ is a bounded sequence of Hermitian $\wdth\times \wdth$ matrices.
Such an operator is a so called tight binding model for a cable with $\wdth$ channels. The terms
$-\Psi_{n+1}\,-\,\Psi_{n-1}$ correspond to the horizontal Laplacian and describe the 'hopping' of an electron from state to state
along the wire. The matrix potentials $V_n$ describe the hopping or interaction between the different channels and may also include some potential.
A particular case of interest are models where the $V_n$ are perturbations of a fixed matrix $W$.
If $V_n=W$ for all $n$ then one finds Bloch waves and the operator describes a pure, space homogeneous cable with a pure crystal structure.
The perturbations then model impurities in the cable.
For instance, randomly doped semiconductors are supposed to be modeled by random potentials $V_n$. 
For instance, the case where $\wdth=1$ and the $V_n$ are independently identically distributed corresponds to 
the one-dimensional Anderson model as proposed by
Anderson \cite{And}.
Choosing $V_n$ to be distributed according to the Gaussian unitary ensemble (GUE) or the
Gaussian orthogonal ensemble (GOE) for $\wdth>1$ corresponds to a Wegner $\wdth$-orbital model.
Wegner\cite{Weg} studied the $\wdth\to\infty$ limit of such models.

The focus in mathematical physics often lies in the spectral theory
on the infinite strip.
From a solid state physics point of view the electronic properties of finite pieces are quite of interest.
The general idea is that absolutely continuous spectrum corresponds to a conductor even for infinitely long pieces, whereas Anderson 
localization corresponds to an isolator when the length of the piece is much larger than the localization length.

In the theory of electronic conduction as developed by Landauer \cite{Lan, Lan2}, Imry \cite{Imr} and B\"uttiker \cite{Buet, Buet2} 
a scattering approach is used. The idea is that the electronic properties of such a finite piece, from now on called the scatterer, 
is in principle given by considering the scattering of
this piece inserted inside an  \emph{ideal lead}. By ideal lead one means a pure cable with only elliptic (propagative) channels.
The mathematical definition will be given in the next section. 
The scattering matrix for this scattering problem describes reflection and transmission of incoming Bloch 
waves to outgoing Bloch waves on the right and left of the scatterer. Related in a twisted way to this scattering matrix 
is the \emph{$S$-transfer matrix} giving the transfer from waves from the left to the right of the scatterer.
We call it $S$-transfer matrix as it is obtained from the scattering matrix and 
we will use the terminology \emph{transfer matrix} for a different object.

From the scattering matrix or the $S$-transfer matrix one can calculate certain quantities such as 
the Landauer conductance or shot noise. For more information on these connections I recommend the review by Beenakker \cite{Been}.
The scattering and the $S$-transfer matrices depend on the specific choice of an ideal lead as well as on 
the choice of a basis for its Bloch waves, but important quantities such as the Landauer conductance do not.

The advantage of the $S$-transfer matrix compared to the scattering matrix 
is the so called multiplicity property. The physics intuition is the following. 
Suppose one puts two scatterers together which are described by $S$-transfer matrices 
$T_1$ and $T_2$. Then the first transfer matrix $T_1$ connects the amplitudes and phase information of waves on the left of scatterer 1 to the
right of scatterer 1 which is the left of scatterer 2. Now, $T_2$ connects these amplitudes and phases to the ones on the right of scatterer 2.
Therefore, the product $T_2 T_1$ connects the amplitudes and phases on the left of the two scatterers to the right of the two scatterers.
Thus, $T_2 T_1$ corresponds to the $S$-transfer matrix of both pieces put together.

\vspace{.2cm}

In the mathematical analysis of operators as given by \eqref{eq-H} one defines the \emph{transfer matrix} from the 
stationary Schr\"odinger equation (cf. \eqref{eq-Tr} and \eqref{eq-Tr-block}). These transfer matrices satisfy the multiplicity property which can be seen easily.
For an ideal lead the transfer matrix is conjugated to a unitary matrix. If one diagonalizes it then it looks like the $S$-transfer matrix
of Bloch waves. In fact, it seems to be quite known that using the same basis change of a disordered piece corresponds to the 
$S$-transfer matrix of this piece w.r.t. the same ideal lead. For instance, this is mentioned in Ref.~\onlinecite{BDR} and it will be confirmed
in this article.

\vspace{.2cm}

An important development in the electronic conduction theory is the so called DMPK\cite{Dor, MePeKu} theory and DMPK equation. 
This is a stochastic differential equation (SDE) describing the conductance of a disordered wire
with respect to its length in a macroscopic setup.
Bachman and de~Roeck\cite{BDR} analyzed the connection of the microscopical Anderson model on a strip to DMPK theory. 
If the unperturbed operator
describes an ideal lead, then they found an SDE describing the evolution of the transfer matrices in an appropriate scaling limit.
This can not be obtained if the unperturbed operator is a pure cable with elliptic (propagative) and hyperbolic (non-propagative) channels.
I believe that in this case one should consider the $S$-transfer matrix coming from scattering a disordered piece 
with respect to the unperturbed operator. 
In these cases the scattering matrices and the $S$-transfer matrices are of lower dimensions than the transfer matrices.
Also, the multiplicity property for the $S$-transfer matrices is no longer valid, but it still holds for the transfer matrices.
The purpose of this paper is to analyze the relations between these matrices in this case (cf. Theorem~\ref{th-main0}).

From a physics point of view, the scattering matrix of a finite disordered piece with respect to a cable with hyperbolic channels does not
only contain information about the scatterer but also about the cable. This is in principle also true if one has an ideal lead, but
since an ideal lead has only propagative channels, it does not affect important quantities such as the Landauer conductance.
However, hyperbolic channels do have an effect. Therefore, it should be treated as a scatterer itself.
By physics intuition, the situation of having the finite scatterer inserted inside an infinite cable should be described by the following limit:
We take a piece of the cable of length $m$, followed
from the finite scatterer and another piece of length $m$ of the cable.,Then we obtain the scattering matrix for these three blocks
together inserted in an ideal lead and take the limit $m\to\infty$ (cf. Figure~\ref{fig1} on page \pageref{fig1}).
We will prove that this limit gives indeed the scattering matrix of the scatterer with respect to the pure cable with 
hyperbolic channels, cf. Theorem~\ref{th-main}.

\vspace{.5cm}

\noindent {\bf Acknowledgment:} I am thankful to H. Schulz-Baldes and A. Klein for many suggestions.

\section{Statement of Results}
\label{sec-model}

As described above, let $H$ be an operator on $\ell^2(\ZZ,\CC^\wdth)\cong\ell^2(\ZZ)\otimes\CC^\wdth$ defined by
\begin{equation} \label{eq-H}
(H \Psi)_n
\,=\,
-\Psi_{n+1}\,-\,\Psi_{n-1}\,+\,V_n \Psi_n\;
\end{equation}
where $V_n\in{\rm Her}(\wdth)$ is a sequence of Hermitian $\wdth\times \wdth$ matrices.
$H$ describes a cable with $\wdth$ channels.
Associated with such an operator are the 
transfer matrices $\Tt^E_n$. They arise from the stationary Schr\"odinger equation
$H\Psi=E\Psi$, which gives
\begin{equation}\label{eq-Tr}
\left(\begin{matrix}
\Psi_{n+1} \\ \Psi_n
\end{matrix}\right)
\,=\,
\Tt^E_n \left( \begin{matrix}
\Psi_{n} \\ \Psi_{n-1} \end{matrix} \right)
\;\qquad\text{for}  \qquad
\Tt^E_n\,=\,
\left(\begin{matrix} V_n - E \one & -\one \\ \one & \nul
\end{matrix} \right)\;.
\end{equation}
Note that $\Tt^E_n$ is in the conjugate
symplectic group ${\rm Sp}(2\wdth)$ defined by
\begin{equation}
{\rm Sp}(2\wdth)=
\left\{ \Tt \in {\rm Mat}(2\wdth,\CC)\,:\,
{\Tt}^* \Jj_\wdth \Tt=\Jj_\wdth\right\}\quad\text{where}\quad
\Jj_\wdth=\begin{pmatrix}
        \nul & \one \\ -\one & \nul
       \end{pmatrix}.
\end{equation}
The individual blocks are all of size $\wdth \times \wdth$.
This group 
is different from the complex symplectic group ${\rm Sp}(2\wdth,\CC)=\{T:T^\top \Jj_\wdth T = \Jj_\wdth\}$.

The transfer matrix of the block of length $\lgth-l$ from $l$ to $\lgth-1$,
where $l<\lgth$, is given by the
product
\begin{equation}\label{eq-Tr-block}
\Tt^E_{l,\lgth}
\,=\,
\Tt^E_{\lgth-1}\,\Tt^E_{\lgth-2}\,\cdots\, \Tt^E_l\quad\text{which gives}\quad
\Tt^E_{l,\lgth} \begin{pmatrix}
                 \Psi_l \\ \Psi_{l-1}
                \end{pmatrix}
=\begin{pmatrix}
  \Psi_\lgth \\ \Psi_{\lgth-1}
 \end{pmatrix}\;
\end{equation}
if $H\Psi=E\Psi$.
This product only depends on $E$ and the sequence $V_l,\ldots,V_{\lgth-1}$. Hence, for a fixed energy $E$, 
each such sequence gives rise to a certain transfer matrix. Moreover, the
transfer matrix for two consecutive blocks (sequences), is just the product of the transfer matrices for each block, e.g.
$\Tt^E_{0,\lgth}=\Tt^E_{0,l}\,\Tt^E_{l,\lgth}$ for $0<l<\lgth$. We referred to this as the multiplicity property in the introduction above.

We want to insert a finite block of length $\lgth$ within a space-homogeneous cable and consider it as a scatterer within the cable.
The scatterer will be described by the sequence $V_0,\ldots,V_{\lgth-1}$ and the transfer matrix $\Tt^E_{0,\lgth}$ which
connects $\binom{\Psi_0}{\Psi_{-1}}$ to $\binom{\Psi_\lgth}{\Psi_{\lgth-1}} $
for a solution of $H\Psi=E\Psi$. 
The space-homogeneous cable will be described by the operator
\begin{equation} \label{eq-H0}
(H_0 \Psi)_n
\,=\,
-\Psi_{n+1}\,-\,\Psi_{n-1}\,+\,W \Psi_n\;,
\qquad
\Psi=(\Psi_n)_{n\in\ZZ}\,\in\,\ell^2(\ZZ,\CC^\wdth)\;.
\end{equation}
The difference to $H$ is that the Hermitian matrix $W$ is always the same and $H_0$ is invariant by translations on $\ZZ$.
Therefore we call $H_0$ space-homogeneous.

Inserting the finite scatterer is described by changing the operator $H_0$ on a finite piece.
Therefore, let $(V_n)_n$ satisfy
\begin{equation} \label{eq-finite-block}
V_n = W \quad\text{for $\;n< 0\;$ and $\;n \geq \lgth\;$},
\end{equation}
then the operator $H$ as defined in \eqref{eq-H} describes the cable with the inserted scatterer given by the sequence
$V_0,\ldots,V_{\lgth-1}$.
The scattering of this piece is described by
the unitary scattering operator $\Ss=\Omega_-^* \Omega_+$ of $H$ with respect to $H_0$,
where $\Omega_{\pm}={\rm s}-\lim_{t\to\mp\infty} e^{\imath tH} e^{-\imath tH_0}$.
Since $\Ss$ commutes with $H_0$, it can be represented by scattering matrices
on the energy shells for (almost) each energy $E$ in the spectrum of $H_0$.

The Hermitian matrix $W$ describes the transverse modes in the cable.
Let $\varphi_\alpha \in \CC^\wdth, \alpha=1,\ldots,\wdth$ be an orthonormal
basis of eigenvectors of $W$ with corresponding real eigenvalues $\lambda_\alpha$.
The spectrum of $H_0$ is purely absolutely continuous and given by the union of $\wdth$ bands,
$\bigcup_{\alpha=1}^\wdth [-2+\lambda_\alpha,2+\lambda_\alpha]$.
Given an energy $E$, $\varphi_\alpha$ is called an elliptic channel if $|\lambda_\alpha-E|<2$, a parabolic channel if $|\lambda_\alpha-E|=2$, 
and a hyperbolic channel if $|\lambda_\alpha-2|>2$. If there is a parabolic channel then $E$ is called a band-edge.
The number of elliptic channels at $E$ will be denoted by $s(E)$, the band-edges are exactly the discontinuities of $s(E)$.
If $E$ is not a band-edge, then the multiplicity of the spectrum of $H_0$ 
at $E$ is given by $2s(E)$ which exactly equals
the number of eigenvalues of modulus $1$ (counted with multiplicity) of the transfer matrix
\begin{equation} \label{eq-freeTr}
\Tt^{0,E}
\,=\,
\left(\begin{matrix}
W-E\one & - \one \\ \one & \nul
\end{matrix} \right)\;.
\end{equation}
Since the multiplicity is $2s(E)$, the scattering matrix describing the scattering operator on the  energy shell
has to be a $2s(E)\times 2s(E)$ matrix. The corresponding extended states of $H_0$ can be split into
$s(E)$ right-moving and $s(E)$ left-moving 
waves at energy $E$. 
In the sequel we will often use $s$ instead of $s(E)$.

\vspace{.2cm}

The terminology \emph{transfer matrix} also appears in
the scattering theory of electronic conduction as developed by Landauer \cite{Lan,Lan2}, Imry \cite{Imr} and 
B\"uttiker \cite{Buet, Buet2}. A short overview is given within a review by Beenakker \cite{Been}.
We will call this transfer matrix the $S$-transfer matrix $\widetilde{\Tt}^E$ in order to distinguish it from the transfer matrix as 
defined above.
The $S$-transfer matrix connects waves on the left to
waves on the right of the finite scatterer, whereas the scattering matrix, let us call it $\Ss^E$, relates incoming and outgoing waves.
For the scattering matrix $\Ss^E$ we choose the following convention. Writing $\Ss^E=\left(\begin{smallmatrix}
                                                                R & T' \\ T & R'
                                                               \end{smallmatrix}\right)$,
the $s \times s$ matrices $T, T'$ correspond to transmission of waves from left to right, resp. right to left, and $R$ and $R'$ correspond 
to reflection of waves on the left, resp. right of the scatterer.
Then, one has the following relations,
\begin{equation}\label{eq-Ss-Tt}
\Ss^E\,
\left(\begin{matrix}
a^+\\b^-
\end{matrix}\right)
\,=\,
\left(\begin{matrix}
a^-\\b^+\end{matrix}\right)
\quad\Leftrightarrow \quad
\widetilde{\Tt}^E\,
\left(\begin{matrix}
a^+\\a^-
\end{matrix}\right)
\,=\,
\left(\begin{matrix}
b^+\\b^-\end{matrix}\right)\;,
\end{equation}
where $a^+, b^+ \in\CC^s$ are vectors describing the amplitudes of right-moving waves on the left, resp. right side of the scatterer, and
$a^-,b^- \in \CC^s$ describe the amplitudes of left-moving waves on the left resp. right side.

As the scattering operator is unitary, the scattering matrices are unitary as well, i.e. $\Ss^E \in {\rm U}(2s)$.
From the relation \eqref{eq-Ss-Tt} one then finds that $\widetilde\Tt^E$ is in the pseudo-unitary or Lorentz group 
${\rm U}(s,s)$ of signature $(s,s)$, defined by
\begin{equation} 
\label{eq-def-U(s,s)}
{\rm U}(s,s)=\left\{\widetilde\Tt\in{\rm Mat}(2s,\CC)\,:\,
\widetilde\Tt^* \Gg_s \widetilde\Tt = \Gg_s\right\}\quad\text{where}\quad
\Gg_s=
\left(\begin{matrix}
          \one & \nul \\ \nul & -\one
         \end{matrix} \right)\;.
\end{equation}
The blocks in $\Gg_s$ are all of size $s\times s$ making it a $2s\times 2s$ matrix.
The conjugate symplectic group ${\rm Sp}(2s)$ and the Lorentz group ${\rm U}(s,s)$ are related by the Cayley matrix,
\begin{equation} \label{eq-def-C_s}
 \Cc_s\,{\rm Sp}(2s)\,\Cc_s^*\,=\,{\rm U}(s,s)\;,\quad\text{where}\quad
\Cc_s=\tfrac{1}{\sqrt{2}}\,\left(\begin{matrix}
                                  \one & i\one \\ \one & -i\one
                                 \end{matrix}\right)\,\inß·{\rm U}(2s)\;.
\end{equation}

As described in Ref.~\onlinecite{MarLan, MePeKu, Been} and in Appendix~\ref{sec-scattering-matrix}, 
$\Ss^E$ and $\widetilde{T}^E$ are related by the polar decompositions
\begin{align}
\label{eq-Tt1}
\widetilde\Tt^E
&\,=\,
\left(\begin{matrix}
U_{r,+} & \nul \\ \nul & U_{r,-} \end{matrix}\right)
\left(\begin{matrix}
\sqrt{Q} & \sqrt{Q-\one} \\ \sqrt{Q-\one} & \sqrt{Q}
\end{matrix}\right)
\left(\begin{matrix}
U_{l,+} & \nul \\ \nul & U_{l,-} \end{matrix}\right) \\
\label{eq-Ss1}
\Ss^E
&\,=\,
\left(\begin{matrix}
 U_{l,-}^* & \nul \\ \nul & U_{r,+} \end{matrix}\right)
\left(\begin{matrix}
-\sqrt{\one-Q^{-1}} & \sqrt{Q^{-1}} \\ 
\sqrt{Q^{-1}} & \sqrt{\one-Q^{-1}} \end{matrix}\right)
\left(\begin{matrix}
U_{l,+} & \nul \\ \nul &  U_{r,-}^*
\end{matrix}\right)\;.
\end{align}
Here $Q$ is a real, diagonal matrix satisfying $Q\geq\one$ and $U_{l,\pm}, U_{r,\pm} \in {\rm U}(s)$ are unitary matrices mixing 
the channels on the left and the right. 
As shown in Appendix~\ref{sec-scattering-matrix}
for any pseudo-unitary matrix $\widetilde\Tt^E$ one finds
a unitary matrix $\Ss^E$ satisfying \eqref{eq-Ss-Tt} by these polar decompositions. 
However, given $\Ss^E$ one may not always find $\widetilde\Tt^E$, as the matrix 
$Q^{-1}\leq\one$ occurring in the polar decomposition of $\Ss^E$ is not necessarily invertible.

We will show that the $S$-transfer matrix and the transfer matrix are related. In fact for energies where $H_0$ has only elliptic channels, 
they are simply related by a conjugation. This is a well known fact and appears e.g. as a Lemma in Ref.~\onlinecite{BDR}.
Using the Lippmann-Schwinger equation, it will be confirmed once more.
The new investigation in this paper is the relation if the background operator $H_0$ has hyperbolic
channels. Then the $S$-transfer matrix is of smaller size than the transfer matrix and the relation between them is more complicated.
More precisely, we obtain the following.
\begin{theo} \label{th-main0}
{\rm (i)} There is a unitary operator $\Uu:\ell^2(\ZZ,\CC^\wdth)\to\int^\oplus \CC^{2s(E)}{\rm d}E$, only depending on $H_0$, and, 
except for finitely many energies in the spectrum of $H_0$, there exists
a scattering matrix $\Ss^E\in{\rm U}(2s(E))$,
such that the spectral decompositions of $H_0$ and the scattering operator $\Ss$ are given by
\begin{equation}
H_0\,=\,
\Uu^* \, \left[\int^\oplus E\,\one_{2s(E)} {\rm d}E
\right]\,\Uu\;,\qquad
\Ss \,=\, \Uu^*\, \Ww\Vv_\lgth \left[\int^\oplus \Ss^E\,{\rm d}E\right] \Vv_\lgth\Uu\,,
\end{equation}
with
\begin{equation}
 \Vv_\lgth\,=\,\int^\oplus \begin{pmatrix}
                                      \one & \nul \\ \nul & e^{-ik_E\lgth} 
                                     \end{pmatrix} {\rm d}E \;,\quad
\Ww\,=\,\int^\oplus \begin{pmatrix}
                       \nul & \one_{s(E)} \\ \one_{s(E)} & \nul
                      \end{pmatrix}{\rm d}E\;.
\end{equation}
Here, $k_E$ is a real, $s(E)\times s(E)$ diagonal matrix and its entries are the wave numbers for the 
extended states of $H_0$ at energy $E$.\\
{\rm (ii)} Let $H_0$ have only elliptic channels at $E$.
Then $\Ss^E$ as in {\rm (i)} and the $S$-transfer matrix $\widetilde\Tt^E$
defined by \eqref{eq-Ss-Tt} both exist. Moreover, there exists $M\in{\rm Sp}(2\wdth)$ {\rm(}defined in \eqref{eq-defM}{\rm)}
only depending on $E$ and $H_0$, such that
\begin{equation}\label{eq-S-tr-ideal}
 \widetilde\Tt^E\,=\,
\Cc_\wdth M^{-1} \Tt^E_{0,\lgth} M \Cc_\wdth^*\;.
\end{equation}
In particular, $\Ss^E$ and $\Tt^E_{0,\lgth}$ are related by the basis change \eqref{eq-S-tr-ideal} and the polar decompositions 
\eqref{eq-Tt1}, \eqref{eq-Ss1}.\\
{\rm (iii)} For all but finitely many energies $E$ where $H_0$ has $s>0$ elliptic and $\wdth-s>0$ hyperbolic channels
there exist $\Ss^E$ as in {\rm(i)} and $\widetilde\Tt^E$ defined by \eqref{eq-Ss-Tt}. 
Moreover, there are matrices $M\in{\rm Sp}(2\wdth)$ {\rm(}defined in \eqref{eq-defM}{\rm)} only depending on $E$
and $H_0$, such that
\begin{equation}\label{eq-S-tr}
 \widetilde\Tt^E\,=\,
\Cc_s O M^{-1} \Tt^E_{0,\lgth} M\left[\one-\hat O^* \left(\hat O M^{-1} \Tt^E_{0,\lgth} M \hat O^* \right)^{-1} \hat O
M^{-1}\Tt^E_{0,\lgth} M \right] O^* \Cc_s^*\;,
\end{equation}
where
$$
O=\left(\begin{smallmatrix}
                              \one & \nul & \nul & \nul \\ \nul & \nul & \one & \nul
                             \end{smallmatrix} \right) \in {\rm Mat}(2s\times2\wdth)
\,\quad\text{and}\quad \hat O=\left(\nul\,\one\,\nul\,\nul \right)\in{\rm Mat}(\wdth-s\times 2\wdth)\;.
$$
The rows are divided in 4 blocks of sizes $s,\wdth-s,s,\wdth-s$ and $\one$ always denotes a unit square matrix.
As the conjugation with $O$ reduces the dimension, we call $\Cc_s^* \widetilde \Tt^E \Cc_s\in{\rm Sp}(2s)$ and also
$\widetilde\Tt^E\in{\rm U}(s,s)$ itself a 'reduced' transfer matrix.
\end{theo}

\noindent {\bf Remarks.}
1. {The unitary operator $\Uu$ is chosen such that the first $s(E)$ entries of $(\Uu\Psi)(E)$ correspond to right moving waves 
and the other ones to left moving waves.
The off diagonal block structure appearing in the direct integral in the definition of $\Ww$ 
interchanges right and left moving waves and is necessary in the
used convention for the scattering matrix $\Ss^E$, as the diagonal blocks correspond to reflection, not transmission.}\\
2. { The expressions $e^{-ik_E\lgth}$ appearing in the definition of $\Vv_\lgth$ correspond to different phase normalizations for waves on the
right and the left of the scatterer.
This way, if $H=H_0$ and hence $\Ss=\one$, then one has
$$
\Ss^E=\begin{pmatrix}
       \nul & e^{ik_E\lgth} \\ e^{ik_E\lgth} & \nul
      \end{pmatrix}\;,\qquad
\widetilde\Tt^E=\begin{pmatrix}
                 e^{ik_E\lgth} & \nul \\ \nul & e^{-ik_E\lgth}
                \end{pmatrix}\;.
$$
Therefore, the $S$-transfer matrix for a piece of length $\lgth$ of the cable $H_0$ 
gives precisely the phase evolution of the waves in that piece. This is a reasonable convention for the $S$-transfer matrix.
}\\
3. {The finitely many energies, where $\widetilde\Tt^E$ does not exist, consist of the
band edges (discontinuities of $s(E)$) and the energies where $\hat O^* M^{-1} \Tt^E_{0,\lgth} M \hat O$ 
as in \eqref{eq-S-tr} is not invertible.
The latter is the case if either $E$ is an eigenvalue of $H$ or if some waves of $H_0$ with energy $E$ are totally reflected. 
We show that these cases 
happen only at finitely many energies $E$.
}

\vspace{.2cm}

In the theory of electronic conduction developed in Ref.~\onlinecite{Buet, Buet2, Imr, Lan, Lan2}, 
the scatterer is connected to so called \emph{ideal leads}. In this case
the $S$-transfer matrices coming from scattering theory are supposed to have the multiplicity property, i.e.
the $S$-transfer matrix for two consecutive blocks is
just the product of the ones for each individual block.
In fact, for energies $E$ where $H_0$ has the maximal possible multiplicity, i.e. $s(E)=\wdth$, 
the $S$-transfer matrix $\widetilde\Tt^E$ is related to the transfer matrix $\Tt^E_{0,\lgth}$ 
by a simple basis change as given in \eqref{eq-S-tr-ideal}.
Since the multiplicity property mentioned above is obviously true
for the transfer matrix, it follows for the $S$-transfer matrix in this case.

However, if $H_0$ has hyperbolic channels, $s(E)< \wdth$, then the $S$-transfer matrix as in \eqref{eq-S-tr}
does not have this property anymore.
For that reason, we make the following definition.
\begin{defini}
 Let $H_I$ be an operator on $\ell^2(\ZZ,\CC^\wdth)$ given by
\begin{equation} \label{eq-HI}
(H_I \psi)_n\,=\,
\,-\, \psi_{n+1} \,-\, \psi_{n-1} \,+\, W_I \psi_n\;,
\end{equation}
where $W_I\in{\rm Her}(\wdth)$ is a Hermitian $\wdth \times \wdth$ matrix.
$H_I$ is called \emph{ideal} at an energy $E$ iff all channels are elliptic for that energy.
Equivalently, this means that the multiplicity of the spectrum of $H_I$ is equal to $2\wdth$ in a neighborhood of $E$.
\end{defini}

\begin{figure}[ht]
\includegraphics[width=0.9\textwidth]{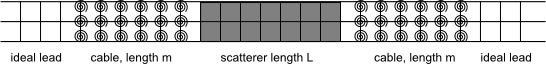} 
\caption{\label{fig1}Scatterer and pieces of cable inserted inside ideal lead.}
\end{figure}

If $H_0$ is not ideal at $E$ then one might consider it as a scatterer with respect to an ideal lead at the energy $E$. 
In particular, from physics intuition, the scattering matrix of the finite block with respect to the non-ideal lead $H_0$ should be described by 
the following limit:
Take a piece of the cable described by $H_0$ of length $m$, followed
from the finite scatterer described by the sequence $V_0,\ldots,V_{\lgth-1}$ and another piece of length $m$ of the cable described by $H_0$,
connect them to an ideal lead on the right and the left, calculate the scattering matrix and take the limit $m\to \infty$
(cf. Figure~\ref{fig1}).

Therefore, let $E$ be an energy where $H_0$ has $s<\wdth$ elliptic channels and where the $S$-transfer matrix and the scattering matrix $\Ss^E$ 
exist. We construct an ideal operator $H_I$ at energy $E$ by defining an appropriate hermitian matrix $W_I$.
The spectral decomposition of $W$ is given by $W=\sum_\alpha \lambda_\alpha \varphi_\alpha \varphi_\alpha^*$.
Assume that the $\varphi_\alpha$ for $\alpha\leq s$ are precisely the elliptic channels, then we define
\begin{equation}\label{eq-def-WI}
W_I\,=\,\sum_{\alpha=1}^s \lambda_\alpha \varphi_\alpha \varphi_\alpha^*\,+\,\sum_{\alpha=s+1}^\wdth E \varphi_\alpha\varphi_\alpha^*\;.
\end{equation}
and let $H_I$ be given by \eqref{eq-HI}.
Furthermore, let the operators $H(m)$ be defined by
\begin{equation}\label{eq-def-H(m)}
(H(m)\Psi)_n\,=\,-\Psi_{n+1}-\Psi_{n-1} + V_n(m) \Psi_n
\end{equation}
with
$V_n(m)=V_n$ for $0\leq n \leq \lgth-1$, $V_n(m)=W$ for $-m\leq n < 0$ and $\lgth\leq n<\lgth+m$ and $V_n(m)=W_I$ for
$n<-m$ and $n\geq \lgth+m$ (cf. Figure~\ref{fig1}).
The scattering of the operators $H(m)$ with respect to $H_I$ at energy $E$ is described 
by scattering matrices $\Ss^E_I(m)$ as in Theorem~\ref{th-main0} (where $H$ gets replaced by $H(m)$, and $H_0$ is replaced by $H_I$). 
$\Ss^E_I(m)$ exists as $H_I$ is ideal at $E$.
\begin{theo} \label{th-main}
Let the $2s\times 2s$ scattering matrix at $E$ for the scattering operator of $H$ with respect to $H_0$ be given by
$\Ss^E=\left(\begin{smallmatrix} R & T' \\ T & R' \end{smallmatrix}\right)$, written in $s\times s$ blocks.
There is a diagonal, unitary
matrix $D$, and there is a real diagonal $\wdth-s \times \wdth-s$ matrix $\theta$,
such that for the scattering matrix $\Ss_I^E(m)$ describing the scattering of $H(m)$ with respect to $H_I$ one has
\begin{equation}\label{eq-SI-lim}
\lim_{m\to\infty} D^m \Ss^E_I(m) D^m\,=\, 
\left(\begin{smallmatrix}
R & \nul & T' & \nul \\
\nul & - e^{\imath\theta} & \nul & \nul \\
T & \nul & R' & \nul \\
\nul & \nul & \nul & e^{\imath\theta}
\end{smallmatrix}\right)\;.
\end{equation}
The whole matrix has size $2\wdth\times 2\wdth$ and is divided in blocks of sizes $s, \wdth-s, s, \wdth-s$.
In particular, using the $2s\times2\wdth$ matrix $O$ as in Theorem~\ref{th-main0}~(iii), we obtain 
\begin{equation}
\Ss^E\,=\, O \left[ \lim_{m\to \infty} D^m \Ss^E_I(m) D^m \right] \,O^*\;.
\end{equation}
\end{theo}
\noindent {\bf Remarks.}
1. {The unitary, diagonal matrices $D^m$ are just phase normalizations counteracting the phase evolution on the pieces of length $m$ from
the cable described by $H_0$.
The $e^{i\theta}$ terms correspond to a total reflection in the hyperbolic channels of $H_0$ in the limit with some specific phase change.}\\
2. {The transfer matrix for the inserted piece in the ideal lead $H_I$, described by $H(m)$, is given by
${(\Tt^{0,E})}^m \Tt^E_{0,N} {(\Tt^{0,E})}^m$ which is related to $\Ss^E_I(m)$ by Theorem~\ref{th-main0}~(ii).
In this sense, the reduced transfer matrix which is related to $\Ss^E$ can be interpreted as some sort of limit of
${(\Tt^{0,E})}^m \Tt^E_{0,N} {(\Tt^{0,E})}^m$ for $m\to\infty$, combined with a projection on the elliptic channels.}

\vspace{0.2cm}

One of the interesting byproducts of this work is the reduced transfer matrix and its relation to the transfer matrix
as given by \eqref{eq-S-tr}. For the reduced transfer matrix, the hyperbolic channels get eliminated in a specific way.
Let me briefly explain with some conjectures why I believe this object is of further interest.

Assume the matrix potentials $V_n$ are random perturbations of $W$, 
i.e. $V_n=W+\lambda W_n$ where $\lambda$ is small and the $W_n$ are i.i.d. random Hermitian matrices with mean zero.
Then the transfer and scattering matrices are random.
If $H_0$ has only elliptic channels, Bachmann and de Roeck \cite{BDR} as well as Valko and Virag \cite{VV} obtained a
stochastic differential equation (SDE) for the evolution 
of the transfer matrix $\Tt^E_{0,\lgth}$ in the limit $\lambda\to 0,\,\lgth=c\lambda^{-2} \to\infty$.
In the presence of hyperbolic channels, such a result can not be obtained.
The main motivation for Bachmann and de Roeck \cite{BDR} was to investigate the relation of such models to DMPK\cite{Dor,MePeKu}  theory 
which studies transport in
disordered wires using scattering matrices. Therefore, the reduced transfer matrix may be of interest.

\vspace{.2cm}

\noindent {\bf Conjecture 1:}
{\it The evolution of the random reduced transfer matrix can be described by an SDE in the appropriate scaling limit 
$\lambda\to 0,\,\lgth=c\lambda^{-2} \to \infty$.} 

\vspace{.2cm}

Related to Ref. \onlinecite{BDR} and \onlinecite{VV} is the perturbative calculation of the invariant measure 
of the random action of the transfer matrices on the flag manifold in the limit $\lambda\to 0$. This action is studied to obtain the
Lyapunov exponents.
For energies where $H_0$ has only elliptic channels, Sadel and Schulz-Baldes \cite{SS} showed under generic conditions on the randomness, 
that the weak-$*$ limit of the invariant measure exists and has a smooth density with respect to a canonical Haar measure.
This weak-$*$ limit distribution could also be obtained from the limit SDE.
In the presence of hyperbolic channels, such a limit distribution should exist and be supported on a certain stable submanifold
determined by the hyperbolic channels, as explained by R\"omer and Schulz-Baldes \cite{RS} who did some numerical calculations.
The stable submanifold is isomorphic to a flag manifold on which the reduced transfer matrices act.
Therefore, I have the following conjecture.

\vspace{.2cm}

\noindent {\bf Conjecture 2:}
{\it  The perturbative invariant measure on the stable submanifold is related to a limit SDE as in Conjecture~1.}

\vspace{.2cm}

Let me give a short outline.
In Section~\ref{sec-channels} we will consider the spectral decomposition of $H_0$ and
scattering states of $H$. In Section~\ref{sec-normal} we obtain some normal forms of the transfer matrix $\Tt^E_{0,\lgth}$ 
after certain basis changes. Guided by physics intuition we define a reduced transfer matrix
in Section~\ref{sec-reduced}. 
In Section~\ref{sec-scattering} we use the Lippmann-Schwinger equations to obtain Theorem~\ref{th-main0}.
Finally, we show Theorem~\ref{th-main} in Section~\ref{sec-ideal}.

\section{Channels and scattering states}
\label{sec-channels}

We will use Dirac notations, hence expressions like  $|\Psi\rangle$ denote vectors 
$\Psi \in (\CC^\wdth)^\ZZ$ (not necessarily in $\ell^2$) and
for $n \in \ZZ$ we denote
the vector $\Psi_n \in \CC^\wdth$ by $\langle n|\Psi \rangle$.
Let $|n,l\rangle$  for $n\in \ZZ, l\in\{1,\ldots,\wdth\}$ denote the $\ell^2(\ZZ,\CC^\wdth)$ vector
defined by $\langle m|n,l\rangle=\delta_{m,n} e_l$, where $e_l$ is the $l$-th canonical basis vector in $\CC^\wdth$.

As above, let $\varphi_\alpha \in \CC^\wdth$, $\alpha=1,\ldots,\wdth$ be an orthonormal basis
of eigenvectors of the Hermitian matrix $W$ and denote the corresponding eigenvalue
by $\lambda_\alpha$, {\it i.e.} 
$W \varphi_\alpha= \lambda_\alpha \varphi_\alpha$.
Furthermore, define 
$|\Psi^0_\alpha,k\rangle$ by 
\begin{equation}\label{eq-P0-k}
\langle n | \Psi^0_\alpha,k\rangle\,=\,\varphi_\alpha 
e^{\imath k n}\,,
\end{equation} 
then one finds
\begin{equation}\label{eq-H_0-P-k}
H_0 |\Psi^0_\alpha,k\rangle = (- 2\cos(k)+\lambda_\alpha) |\Psi^0_\alpha,k\rangle\;.
\end{equation}
These pseudo-eigenvectors form a partition of unity in the sense that
$$
\sum_{\alpha=1}^\wdth \frac{1}{2\pi} \int_{-\pi}^\pi
\,\langle m,j |\Psi^0_\alpha,k\rangle \langle \Psi^0_\alpha,k| n,l \rangle \,{\rm d}k
\,=\, \delta_{m,n} \delta_{j,l}\,=\, \langle m,j | n,l\rangle
$$
Therefore, in a weak operator topology induced by the functionals $B\mapsto \langle m,j | B | n,l\rangle$ 
(as I am not testing with all $\ell^2$ vectors this topology is actually weaker than the usual weak operator topology)
one can write
\begin{equation}\label{eq-part1}
\one 
\,=\,
\sum_{\alpha=1}^\wdth \frac{1}{2\pi} \int_{-\pi}^\pi
\,|\Psi^0_\alpha,k\rangle \langle \Psi^0_\alpha,k| \,{\rm d}k\;.
\end{equation}
Extending the Fourier transform $L^2((-\pi,\pi),\frac{{\rm d}k}{2\pi}) \rightarrow \ell^2(\ZZ)$ to distributions, 
one formally obtains from the inverse transform
\begin{equation} \label{eq-delta}
 \frac{1}{2\pi} \langle \Psi^0_\beta,k' | \Psi^0_\alpha,k\rangle\,=\,
\sum_{n\in\ZZ} \frac{1}{2\pi}\, \varphi_\beta^* \varphi_\alpha\, e^{i(k-k')n}\,=\,\delta_{\alpha,\beta}\,\delta(k-k')\;.
\end{equation}
In this sense, the pseudo-eigenvectors $|\Psi^0_\alpha,k\rangle$ form an orthogonal system.

Recall that we called an eigenvector $\varphi_\alpha$ of $W$ an elliptic channel for the energy $E$
iff $|E-\lambda_\alpha|<2$. In that case there exists $k_\alpha\in (0,\pi)$ such that
\begin{equation} \label{eq-E-k}
E=-2\cos(k_\alpha)+\lambda_\alpha\;.
\end{equation}
The terminology \emph{elliptic} comes from the fact,
that this corresponds to eigenvalues $e^{\pm\imath k_\alpha}$ of the transfer matrix $\Tt^{0,E}$ and is
therefore related to a rotation.
Now consider $k_\alpha$ as a function 
$k_\alpha(E)$, where the interval on which this function is defined depends on $\alpha$.
To change the normalization of the pseudo-eigenvectors with respect to  energy, define
for the elliptic channels
\begin{equation} \label{eq-P0-E}
|\Psi^0_\alpha,E,\pm\rangle
\,=\,
(4\pi\sin(k_\alpha))^{-1/2} |\Psi^0_\alpha,\pm k_\alpha\rangle\;
\end{equation}
which by \eqref{eq-H_0-P-k} and \eqref{eq-E-k} are pseudo-eigenvectors of $H_0$ with energy $E$.
A change of variables in \eqref{eq-part1} shows
\begin{equation}
\label{eq-part1E}
\one
\,=\,
\sum_{\alpha=1}^\wdth
\int_{-2+\lambda_\alpha}^{2+\lambda_\alpha}
 \left(|\Psi^0_\alpha,E,+\rangle\langle\Psi^0_\alpha,E,+|
\;+\;|\Psi^0_\alpha,E,-\rangle\langle\Psi^0_\alpha,E,-|\right){\rm d}E\;,
\end{equation}
and the spectral decomposition of $H_0$ is given by
\begin{equation}\label{eq-sp-H_0}
H_0\,=\,
\sum_{\alpha=1}^\wdth
\int_{-2+\lambda_\alpha}^{2+\lambda_\alpha}
E\,\left(|\Psi^0_\alpha,E,+\rangle\langle\Psi^0_\alpha,E,+|
\;+\;|\Psi^0_\alpha,E,-\rangle\langle\Psi^0_\alpha,E,-|\right) {\rm d}E\;.
\end{equation}

Furthermore, we say that
$\varphi_\alpha$ is an hyperbolic channel iff $|E-\lambda_\alpha|>2$
and a parabolic channel iff $|E-\lambda_\alpha|=2$. The parabolic channels correspond to band edges and there are at most
$2\wdth$ of them.
Now let $E$ be some energy in the spectrum of $H_0$
without any parabolic channel.
Then there is at least one elliptic channel for $E$.
Let us reorder the channels such that $\varphi_1,\ldots,\varphi_s$ are elliptic 
and $\varphi_{s+1},\ldots,\varphi_\wdth$ are hyperbolic channels.
Furthermore, for the hyperbolic channels $\alpha>s$ define $\gamma_\alpha>0$ 
and $u_\alpha \in\{-1,1\}$  such that
\begin{equation}\label{eq-def-gm}
E
\,=\,
- 2u_\alpha \cosh(\gamma_\alpha)\,+\,\lambda^\alpha\;, \quad (\alpha > s)\;.
\end{equation}
Then the $2\wdth$ vectors
$$
\left(\begin{matrix}
\varphi_\alpha \\ e^{\pm \imath k_\alpha} \varphi_\alpha
\end{matrix} \right)
\;, 1\leq \alpha \leq s\;,\quad\text{and}\quad
\left(\begin{matrix}
\varphi_\alpha \\
u_\alpha e^{\pm\gamma_\alpha} \varphi_\alpha
\end{matrix}\right)\, s<\alpha\leq \wdth
$$
are eigenvectors of $\Tt^{0,E}$ and form a basis of $\CC^{2\wdth}$.
Any formal eigenvector $\Psi^0$ of $H_0$ satisfying
$H_0 \Psi^0=E\Psi^0$ is uniquely defined by $\Psi^0_0$ and $\Psi^0_1$ and hence a linear combination of the $2\wdth$
formal eigenvectors
$|\Psi^0_\alpha,E,+\rangle$, $|\Psi^0_\alpha,E,-\rangle$ ($\alpha \leq s$)
and $|\hat\Psi^0_\alpha,E,+ \rangle$, $|\hat\Psi^0_\alpha,E,- \rangle$ ($\alpha>s$) given by
\begin{equation}\label{eq-hP0-E}
\langle n | \hat\Psi^0_\alpha,E,\ominus \rangle=\varphi_\alpha 
[2\pi\sinh(\gamma_\alpha)]^{-\frac12}
\,u_{\alpha}^{\delta_{\ominus,+}}\, u_\alpha^{n}\,e^{\ominus \gamma_\alpha n}\,,\quad \alpha>s\,, \ominus\in\{+,-\}.
\end{equation}
Here and below we use $\ominus$ as variable symbol for $+$ or $-$. 
In this sense, $\delta_{\ominus,+}=1$ for $\ominus=+$ and $\delta_{\ominus,+}=0$ for $\ominus=-$. For a number $c$ we define
$\ominus c=\delta_{\ominus,+} c - \delta_{\ominus,-} c$.
The factor in \eqref{eq-hP0-E} seems strange but it leads to nice relations in the next section.

Thus, for a formal eigenvector $|\Psi^0,E\rangle$ of $H^0$  there are
coefficients $c_\alpha^+,c_\alpha^-$ for $\alpha\leq s$
and $\hat{c}_\alpha^+, \hat{c}_\alpha^-$ for $\alpha >s$ such that
$$
\langle n |\Psi^0, E\rangle
\,=\,
\sum_{\alpha\leq s, \ominus \in\{+,-\}}
c^\ominus_\alpha |\Psi^0_\alpha,E,\ominus\rangle\;+\;
\sum_{\alpha>s,\ominus\in\{+,-\}} \hat{c}^\ominus_\alpha
|\hat\Psi^0_\alpha,E,\ominus\rangle
$$
Now let $|\Psi,E\rangle$ be some 
formal eigenvector of $H$ with eigenvalue $E$.
Then for $n\leq 0$ and $n \geq \lgth-1$ it looks like a formal 
eigenvector of $H_0$. Therefore, there are constants
$a^+_\alpha, a^-_\alpha, b^+_\alpha, b^-_\alpha$ for $\alpha\leq s$
and $\hat{a}^+_\alpha, \hat{a}^-_\alpha, \hat{b}^+_\alpha, 
\hat{b}^-_\alpha$ for $\alpha>s$ associated to $|\Psi,E\rangle$ by
\begin{align}
\langle n|\Psi,E\rangle
&=
\sum_{\substack{\alpha\leq s \\ \ominus\in\{+,-\}}}
a^\ominus_\alpha \langle n|\Psi^0_\alpha,E,\ominus\rangle
\;+\;\sum_{\substack{\alpha>s \\ \ominus\in\{+,-\}} }
\hat{a}^\ominus_\alpha \langle n | \hat\Psi^0_\alpha,E,\ominus \rangle \;,\quad
\label{eq-scst1}
\end{align}
for $n\leq 0$ and
\begin{align} 
\langle n|\Psi,E\rangle
&=
\sum_{\substack{\alpha\leq s \\ \ominus\in\{+,-\}}}
b^\ominus_\alpha\,e^{-\ominus\imath k_\alpha \lgth} \langle n|\Psi^0_\alpha,E,\ominus\rangle
+\sum_{\substack{\alpha>s \\ \ominus\in\{+,-\}}}
\hat{b}^\ominus_\alpha \,e^{-\ominus\gamma_\alpha \lgth} \langle n | \hat\Psi^0_\alpha,E,\ominus \rangle
\label{eq-scst2}
\end{align}
for $n\geq \lgth-1$.
$|\Psi,E\rangle$ is an eigenvector of $H$ iff there are only exponential 
decaying parts for the limits $n\to \pm \infty$, which means that
$a^+=a^-=b^+=b^-=0,\, \hat{a}^-=\hat{b}^+=0$ where $a^+, \hat{a}^+$ 
denote the vectors $(a^+_\alpha)_{1\leq\alpha\leq s}$,
$(\hat{a}^+_\alpha)_{s<\alpha\leq \wdth}$ and $a^-, \hat{a}^-,b^+, b^-, \hat b^+, \hat b^-$
are correspondingly defined.
$|\Psi, E\rangle$ is called a scattering state, extended state, or pseudo-eigenvector of $H$
iff it is not an eigenvector and has no exponential growing parts, neither at $+\infty$ nor at $-\infty$,
which means $\hat{a}^-=\hat{b}^+=0$.
(These are the states that can be used to create a sequence $|\Psi_n\rangle$ of normalized $\ell^2$ vectors by cut offs, such that
$\left\|(H-E)|\Psi_n\rangle\right\|\to 0 $ for $n\to \infty$. Hence by the Weyl criterion, $E$ is in the spectrum of $H$ if a scattering
state exists.)
Thus, a pseudo-eigenvector $|\Psi,E\rangle$ includes at least one 
elliptic channel on at least one side.
Therefore $\langle n |\Psi,E\rangle$ is not going to zero for $n\to \infty$ or $n\to -\infty$ but 
$\langle n |\Psi,E\rangle$ is bounded.

\section{Normal forms of the transfer matrices}
\label{sec-normal}

For a formal eigenvector $|\Psi,E\rangle$ of $H$,
the coefficients are related by the transfer 
matrix and one has
$\Tt^E_{0,\lgth}
\left(\begin{smallmatrix}
\langle 0 | \Psi,E\rangle \\
\langle -1| \Psi,E\rangle
\end{smallmatrix}\right)
=
\left(\begin{smallmatrix}
\langle \lgth|\Psi,E\rangle \\
\langle \lgth-1 | \Psi,E\rangle
\end{smallmatrix}\right)\;.
$
Using the notations as in \eqref{eq-scst1}  and \eqref{eq-scst2} one obtains from
\eqref{eq-P0-k}, \eqref{eq-P0-E} and \eqref{eq-hP0-E} that
\begin{align*}
\Tt^E_{0,\lgth}&\left[
\sum_{\substack{\alpha\leq s\\ \ominus\in\{+,-\}}}\!\!\!
\frac{a_\alpha^{\ominus}}{(2\sin(k_\alpha))^{1/2}}
\left(\begin{matrix}
\varphi_\alpha \\ \varphi_\alpha e^{-\ominus \imath k_\alpha}
\end{matrix}\right)
+\!\!\!\sum_{\substack{\alpha>s\\ \ominus\in\{+,-\}}} \!\!
\frac{\hat{a}_\alpha^{\ominus}\,(u_\alpha)^{\delta_{+,\ominus}}}{\sqrt{\sinh(\gamma)}}
\left( \begin{matrix}
\varphi_\alpha \\ \varphi_\alpha
u_\alpha e^{-\ominus \gamma_\alpha}
\end{matrix} \right)
\right]
\;=\\
&
\left[
\sum_{\substack{\alpha\leq s\\\ominus\in\{+,-\}}}\!\!\!
\frac{b_\alpha^{\ominus}}{(2\sin(k_\alpha))^{1/2}}
\left(\begin{matrix}
\varphi_\alpha \\ \varphi_\alpha e^{-\ominus \imath k_\alpha}
\end{matrix}\right)
+\!\!\!\sum_{\substack{\alpha>s\\ \ominus\in\{+,-\}}} \!\!\!
\frac{\hat{b}_\alpha^{\ominus}\;(u_\alpha)^{\delta_{+,\ominus}}}{\sqrt{\sinh(\gamma)}}
\left( \begin{matrix}
\varphi_\alpha \\ \varphi_\alpha
u_\alpha e^{-\ominus \gamma_\alpha}
\end{matrix} \right)
\right].
\end{align*}
Working in the conjugate symplectic group one can diagonalize
the hyperbolic channels. In order to do this we define 
\begin{align}
& U=(\varphi_1,\ldots,\varphi_\wdth)\;\in\;{\rm U}(\wdth)\,,  \label{eq-U} \\
& k=\diag(k_1,\ldots,k_s),\quad
\gamma=\diag(\gamma_{s+1},\ldots,\gamma_\wdth),\quad
u=\diag(u_{s+1},\ldots,u_\wdth), \label{eq-k-gm-u}
 \end{align}
and the conjugate symplectic $2\wdth\times 2\wdth$ matrix
\begin{align}\label{eq-defM}
& M
\,=\,
 \left(\begin{matrix} 
U & \nul \\ \nul & U
\end{matrix}\right)
\left(\begin{smallmatrix}
(\sin(k))^{-\frac12} & \nul & \nul & \nul \\
\nul & u(2\sinh(\gamma))^{-\frac12} & \nul & (2\sinh(\gamma))^{-\frac12} \\
\cos(k)(\sin(k))^{-\frac12} & \nul & (\sin(k))^{\frac12} & \nul \\
\nul & e^{-\gamma} (2\sinh(\gamma))^{-\frac12}  & \nul & u e^{\gamma} (2\sinh(\gamma))^{-\frac12} 
\end{smallmatrix}\right)\;.
\end{align}
Then $M$ transforms the free transfer matrix to its symplectic
normal form
\begin{equation}\label{eq-freeTr-normal}
M^{-1} \Tt^{0,E}\,M
\,=\,
\left(\begin{smallmatrix}
\cos(k) & \nul & -\sin(k) & \nul \\
\nul & ue^{\gamma} & \nul & \nul \\
\sin(k) & \nul & \cos(k) & \nul \\
\nul & \nul & \nul & u e^{-\gamma}
\end{smallmatrix} \right)\;,
\end{equation}
and one obtains
\begin{equation}\label{eq-Tr-rel}
\Tt^E
\left( \begin{smallmatrix}
a^+ + a^- \\
2\hat{a}^+ \\
\imath (a^- - a^+) \\ 
2\hat{a}^-
\end{smallmatrix}\right)
\,=\,
\left( \begin{smallmatrix}
b^+ + b^- \\
2\hat{b}^+ \\
\imath(b^- - b^+)\\ 
2\hat{b}^-
\end{smallmatrix}\right)\;,\quad
\text{where}\quad
\Tt^E=M^{-1} \Tt^E_{0,\lgth} M\,.
\end{equation}
Note that $\Tt^E \in{\rm Sp}(2\wdth,\CC)$. 
To diagonalize the elliptic channels for $\Tt^{0,E}$ we need to conjugate $M^{-1} \Tt^{0,E} M$ by the
Cayley matrix as defined in \eqref{eq-def-C_s}. This way we obtain the normal form of the free transfer matrix in the Lorentz group
${\rm U}(\wdth,\wdth)$,
$$
\Cc_\wdth M^{-1} \Tt^{0,E} M \Cc_\wdth^*
\,=\,
\left( \begin{smallmatrix}
e^{\imath k} & \nul & \nul & \nul \\
\nul & u \cosh(\gamma) & \nul & u \sinh(\gamma) \\
\nul & \nul & e^{-\imath k} & \nul \\
\nul & u \sinh(\gamma) & \nul & u \cosh(\gamma)
\end{smallmatrix} \right)\;.
$$
Furthermore one obtains for $\Cc_\wdth \Tt^E \Cc_\wdth^*\,\in\,{\rm U}(\wdth,\wdth)$ that
\begin{equation}\label{eq-Tr-rel2}
\Cc_\wdth \Tt^E \Cc_\wdth^*\; \left( \begin{smallmatrix}
a^+ \\ \hat{a}^+ +\imath \hat{a}^- \\
a^- \\ \hat{a}^+ -\imath \hat{a}^-
\end{smallmatrix} \right)
\,=\,
\left( \begin{smallmatrix}
b^+ \\ {\hat{b}^+ +\imath \hat{b}^-} \\
b^- \\ {\hat{b}^+ -\imath \hat{b}^-} 
\end{smallmatrix} \right)\;.
\end{equation}


\section{Reduced transfer matrix}
\label{sec-reduced}

We want to define a reduced transfer matrix relating
the coefficients for the elliptic channels appearing in
scattering states. This means we look for solutions of the
equations above where $\hat{a}^-=\hat{b}^+=0$.
Given $a^+, a^-$ and $\hat{a}^-=0$ the question is whether there exists a unique
$\hat{a}^+$ such that $\hat{b}^+=0$.
This is the case if the following $(\wdth-s)\times(\wdth-s)$ matrix
\begin{equation}\label{eq-condred}
A_E\,=\,
\begin{matrix} \big(\,
 \begin{smallmatrix}
\nul_{(\wdth-s)\times s} & \one_{(\wdth-s)\times(\wdth-s)}
& \nul_{(\wdth-s)\times s} & \nul_{(\wdth-s)\times (\wdth-s)} \,\big) \end{smallmatrix} \\
& 
\end{matrix}
\Tt^E\;
\left(
\begin{smallmatrix}
\nul_{s\times(\wdth-s)} \\ \one_{(\wdth-s)\times(\wdth-s)} \\ \nul_{s\times(\wdth-s)}
\\ \nul_{(\wdth-s)\times(\wdth-s)}
\end{smallmatrix}\right)
\end{equation}
is invertible. The indices indicate the size of the matrices.
\begin{lemma}
 The matrix $A_E$ is invertible for all but finitely many energies $E$ in the spectrum of $H_0$.
\end{lemma}
\noindent {\bf Proof.}
Let $I$ be a bounded energy interval without parabolic channels. In $I$ the elliptic and hyperbolic channels
as well as the matrices $U$ and u (as defined in \eqref{eq-U} and \eqref{eq-k-gm-u}) stay the same.
Now $\Tt^E_{0,\lgth}$ is of the form
$$
\Tt^E_{0,\lgth}=\begin{pmatrix}
             E^\lgth & \nul \\ \nul & \nul
            \end{pmatrix}\,+\,P(E)
$$
where $P(E)$ is a polynomial in $E$ of degree $\lgth-1$. Hence, we obtain from
\eqref{eq-defM}, \eqref{eq-Tr-rel} and \eqref{eq-condred} for $E\in I$ that
\begin{equation} \label{eq-AE}
 \sqrt{2\sinh(\gamma)} e^{-\gamma} \,A_E\,  \sqrt{2\sinh(\gamma)}\,=\,
E^\lgth + P_1(E) e^{-\gamma}+e^{-\gamma} P_2(E) + e^{-\gamma} P_3(E) e^{-\gamma}\,
\end{equation}
where $P_1(E),\, P_2(E),\,P_3(E)\in{\rm Mat}(\wdth-s,\CC)$ are all polynomials in $E$ of degree $\lgth-1$.
Letting $\Lambda=\diag(\lambda_{s+1},\ldots,\lambda_\wdth)$ one obtains from \eqref{eq-def-gm} that
\begin{equation}
 e^{-\gamma}=e^{-\gamma(E)}=\tfrac12\left[u(\Lambda-E)-\sqrt{[u(\Lambda-E)]^2-4} \right]
\end{equation}
Since $u(\Lambda-E)=2\cosh(\gamma)>2$ for $E\in I$, the functions
$E\mapsto e^{-\gamma(E)}$ and $E\mapsto A_E$ can be extended to complex analytic functions on the strip $I\times \imath\,\RR\subset\CC$.
If the imaginary part $\Im(E)$ tends to $\infty$, then $e^{-\gamma}$ tends to zero. Multiplying $\eqref{eq-AE}$ by $E^{-\lgth}$
and letting $\Im(E)\to\infty$, the right hand side converges to $\one$.
Therefore, $A_E$ is invertible for large $\Im(E)$ and
$\det(A_E)$ is analytic and not identical to zero. Hence,
$\det(A_E)\neq 0$ except for finitely many energies $E$ in the
precompact interval $I$.
As there are only finitely many energies with parabolic channels, this shows the claim.
\hfill $\Box$

\vspace{.2cm}

\noindent {\bf Remark.} 
{
An interesting question might be the meaning if $A_E$ is not invertible.
In this case $A_E$ has a kernel and one can find $\hat a^+$ such that $A_E \hat a^+=0$.
This means 
$$
\Tt^E \left(\begin{smallmatrix}
             0 \\ 2\hat a^+ \\ 0 \\ 0
            \end{smallmatrix} \right)
\,=\,
\left(\begin{smallmatrix}
             b^+ + b^- \\ 0 \\ i(b^- - b^+) \\ 2\hat b^-
            \end{smallmatrix} \right)\;.
$$
If one finds furthermore that $b^+=0$ and $b^- =0$, then this corresponds to an eigenvector of $H$ described by $\hat a^+$ and $\hat b^-$
and $E$ is an eigenvalue of $H$.
If the latter is not the case then we find a scattering state that has no elliptic channel on the left since $a^+=a^-=0$.
From the interpretation of scattering
states which will be given by the Lippmann Schwinger equation this means that there is an extended state or wave which is totally reflected. 
As we have seen, this happens only for finitely many energies.
In particular, $H$ has only finitely many eigenvalues.
}

\vspace{.2cm}

Let us now consider an energy $E$ where $A_E$ is invertible.
Then any vectors $a^+,a^-$ define a unique scattering state
characterized by the coefficients $a^+,a^-,b^+,b^-$ and
$\hat{a}^+,\hat{b}^-$ as defined in \eqref{eq-scst1} and \eqref{eq-scst2}.
More precisely, choosing vectors $a^+, a^-$ and letting
$$
2\hat{a}^+
\,=\,
\;-\;
A_E^{-1}
\;\begin{matrix} 
\big(\begin{smallmatrix} \nul & \one & \nul & \nul \end{smallmatrix} \big) \\
& 
\end{matrix}
\!\!
\Tt^E\;
\left(\begin{smallmatrix}
{a^++a^-} \\ 0 \\ 
\imath(a^--a^+) \\ 0 
\end{smallmatrix}\right)\;
$$
one obtains
$$
\Tt^E \left(\begin{smallmatrix}
a^++a^- \\ 2\hat{a}^+ \\ \imath(a^--a^+) \\ 0
\end{smallmatrix}\right)
\,=\,
\left(\begin{smallmatrix}
b^++b^- \\ 0 \\ \imath(b^--b^+) \\ 2\hat{b}^-
\end{smallmatrix}\right)\;,
$$
and has found all coefficients for the scattering state.
In this case we define the reduced 
$2s \times 2s$ transfer matrix $\hat\Tt^{E}$ by
\begin{equation}\label{eq-redtr}
\hat\Tt^E\,\left(\begin{matrix}
a^++a^- \\ \imath(a^--a^+)
\end{matrix}\right)
\,=\,
\left(\begin{matrix}
b^++b^- \\ \imath(b^--b^+)
\end{matrix}\right)\;.
\end{equation}
Another way to write $\hat\Tt^E$ would be
\begin{equation} \label{eq-redtr0}
\hat\Tt^E
=
\left(\begin{smallmatrix}
\one & \nul & \nul & \nul \\ \nul & \nul & \one & \nul
\end{smallmatrix}\right)\,
\Tt^E\left\{
\one -\left(\begin{smallmatrix} \nul \\ \one \\ \nul \\ \nul
\end{smallmatrix}\right)\;
A_E^{-1}\;
\left(\begin{smallmatrix} \nul \\ \one \\ \nul \\ \nul \end{smallmatrix}\right)^*
\Tt^E \right\}
\left(\begin{smallmatrix}
\one & \nul \\ \nul & \nul \\ \nul & \one \\ \nul & \nul
\end{smallmatrix}\right)\,.
\end{equation}
The size of the first matrix on the right hand side of the equation
is $2s\times 2\wdth$, the columns are divided into two blocks,
each of size $s$, and the rows are divided in
4 blocks of sizes $s,\, \wdth-s,\, s$ and $ \wdth-s$ in that order. This matrix is the same as the matrix $O$ in Theorem~\ref{th-main0}~(iii).
The last matrix is the transpose of the first one. The matrix to the left and right of $A_E^{-1}$ is the same one that appears in 
\eqref{eq-condred} and it is equal to $\hat O^*$ as in Theorem~\ref{th-main0}~(iii).
A conjugation of \eqref{eq-redtr} by the Cayley matrix yields
\begin{equation} \label{eq-redtr2}
 \widetilde\Tt^E \begin{pmatrix}
              a^+ \\ a^-
             \end{pmatrix}
\,=\, \begin{pmatrix}
       b^+ \\ b^-
      \end{pmatrix}\;,\quad
\text{for} \quad \widetilde\Tt^E=\Cc_s \hat\Tt^E \Cc_s^*\;.
\end{equation}

Note, if $s=\wdth$ then there is no hyperbolic channel and therefore
all formal eigenvectors of $H$ are scattering states and $\Tt^E$
already relates the elliptic channels. Therefore, in this case one
simply defines $\hat\Tt^E=\Tt^E$. Then  $\widetilde\Tt^E = \Cc_N \Tt^E \Cc_n^*$ and 
equations \eqref{eq-Tr-rel} and \eqref{eq-redtr} as well as \eqref{eq-Tr-rel2} and \eqref{eq-redtr2} are the same.
In particular, $\hat\Tt^E$ is conjugate symplectic and $\widetilde\Tt^E$ is pseudo-unitary.
This is actually always true, if the reduced transfer matrix exists.
\begin{proposi}
The reduced transfer matrix is conjugate symplectic, i.e.
$\hat\Tt^E\,\in\,{\rm Sp}(2s)$, and consequently, $\widetilde \Tt^E\,\in\,{\rm U}(s,s)$.
\end{proposi}
\noindent {\bf Proof.}
Let $x_i, y_i \in\CC^s,\; i=1,2$ and define
$\hat{x}_i, \hat{y}_i$ by
$$
\hat\Tt^E\left(\begin{matrix} x_i \\ y_i \end{matrix}\right)
\,=\,
\left(\begin{matrix} \hat{x}_i \\ \hat{y}_i \end{matrix} \right)
\;,\;\;\text{for}\;\; i=1,2\,,\quad
\text{then}\quad\exists \;\hat{a}_i,\hat{b}_i\,\in\,\CC^{\wdth-s}\,
\;:\;
\Tt^E \left(\begin{smallmatrix} x_i \\ \hat{a}_i \\ y_i \\ 0
\end{smallmatrix}\right)
\,=\,
\left(\begin{smallmatrix}
\hat{x}_i \\ 0 \\ \hat{y}_i \\ \hat{b}_i
\end{smallmatrix}\right)
$$
and one obtains
\begin{align*}
&\left(\begin{matrix}
x_1 \\ y_1 \end{matrix}\right)^*
\,(\hat\Tt^E)^*\,\Jj_s \,\hat\Tt^E\,
\left( \begin{matrix}
x_2 \\ y_2 \end{matrix}\right)
\,=\,
\left(\begin{matrix}
\hat{x}_1 \\ \hat{y}_1 \end{matrix}\right)^*
\,\Jj_s \,
\left( \begin{matrix}
\hat{x}_2 \\ \hat{y}_2 \end{matrix}\right)
\,=\,
\hat{x}_1^* \hat{y}_2\,-\,\hat{y}_1^* \hat{x}_2 \\
&\quad =\;
\left( \begin{smallmatrix}
\hat{x}_1 \\ 0 \\ \hat{y}_1 \\ \hat{b}_1
\end{smallmatrix}\right)^*
\,\Jj_\wdth\,
\left( \begin{smallmatrix}
\hat{x}_2 \\ 0 \\ \hat{y}_2 \\ \hat{b}_2
\end{smallmatrix}\right)
\,=\,
\left( \begin{smallmatrix}
x_1 \\ \hat{a}_1 \\ y_1 \\ 0
\end{smallmatrix}\right)^*
\,(\Tt^E)^*\Jj_\wdth \Tt^E\,
\left( \begin{smallmatrix}
x_2 \\ \hat{a}_2 \\ y_2 \\ 0
\end{smallmatrix}\right)
\,=\,
\left( \begin{smallmatrix}
x_1 \\ \hat{a}_1 \\ y_1 \\ 0
\end{smallmatrix}\right)^*
\,\Jj_\wdth\,
\left( \begin{smallmatrix}
x_2 \\ \hat{a}_2 \\ y_2 \\ 0
\end{smallmatrix}\right)
\\
&\quad=\;
x_1^* y_2\,-\, y_1^* x_2\,=\,
\left(\begin{matrix}
x_1 \\ y_1 \end{matrix}\right)^*
\;\Jj_s\;
\left(\begin{matrix}
x_2 \\ y_2 \end{matrix} \right)\;.
\end{align*}
As this is true for arbitrary $x_i, y_i$ one has
$(\hat\Tt^E)^*\,\Jj_s\,\hat\Tt^E\,=\,\Jj_s$ and hence
$\hat\Tt^E$ is conjugate symplectic.
\hfill $\Box$

\section{Scattering operator and scattering matrix}
\label{sec-scattering}

Let $\Omega_{\pm}={\rm s-}\lim_{t\to \mp \infty} e^{\imath tH} e^{- \imath t H_0}$ 
be the M\o ller operators and 
$\Ss=\Omega_{-}^* \Omega_+$ the scattering operator.
$\Ss$ commutes with $H_0$ and can therefore be defined as operator on the energy shells
for almost all energies $E$ in the spectrum of $H_0$.
So let
$|\Psi^0_{\text{in}},E\rangle$ be some pseudo-eigenvector of $H_0$
with such an energy $E$, then $|\Psi^0_{\text{out}},E\rangle=\Ss|\Psi^0_{\text{in}},E\rangle$ is defined and
also a pseudo-eigenvector of $H_0$ with the same energy $E$.
The subscripts 'in' and 'out' correspond to the physics intuition that the scattering operator maps the incoming states to the outgoing
states.
Furthermore, for almost all energies $E$ the M\o ller operators can be defined as maps from the energy shell with energy 
$E$ with respect to  the operator $H_0$ ,
to the energy shell with the same energy $E$, with respect to the operator $H$.
In this sense, we have
\begin{equation}\label{eq-in-out}
|\Psi,E\rangle=\Omega^+|\Psi^0_{\text{in}},E\rangle=
\Omega^- |\Psi_{\text{out}}^0,E\rangle\;,\qquad 
|\Psi^0_{\text{out}},E\rangle=
\Ss |\Psi^0_{\text{in}},E\rangle\;,
\end{equation}
where $|\Psi,E\rangle$ is pseudo-eigenvector of $H$.
One obtains from the Lippmann-Schwinger equations \cite{LiSch,ReedSim3},
\begin{align}
\langle n |\Psi,E\rangle
&=
\langle n|\Psi^0_{\text{in}},E\rangle\,\,+\,
\lim_{\epsilon \searrow 0}\,\langle n|(E-H_0-\imath \epsilon)^{-1} (H-H_0) |\Psi,E\rangle,
\label{eq-LipSch1}\\
\langle n |\Psi,E\rangle
&=
\langle n|\Psi^0_{\text{out}},E\rangle\,+\,
\lim_{\epsilon \nearrow 0}\,\langle n|(E-H_0-\imath \epsilon)^{-1} (H-H_0) |\Psi,E\rangle\;.
\label{eq-LipSch2}
\end{align}
Inserting the
partition of unity \eqref{eq-part1} and changing to an integral over the unit circle in the complex plane
by substituting $e^{ik}=z,\; dk=i^{-1}z^{-1}\, dz$ gives
\begin{align*}
& \langle n|(E+H_0-\imath \epsilon)^{-1} (H-H_0) |\Psi,E\rangle \\
& \quad =
\sum_{\alpha=1}^\wdth\frac{1}{2\pi}\,\int_{-\pi}^\pi {\rm d}k\,\left\{
\langle n|\Psi^0_\alpha,k\rangle
\langle\Psi^0_\alpha,k|(E-H_0-\imath \epsilon)^{-1} (H-H_0) |\Psi,E\rangle\right\} \\
& \quad =
\sum_{\alpha=1}^\wdth \int_{-\pi}^\pi\frac{{\rm d}k}{2\pi}\left\{
\frac{\varphi_\alpha e^{\imath k n}}{E+\imath \epsilon+e^{\imath k}+e^{-\imath k}-\lambda_\alpha}
\left[\sum_{m=0}^{\lgth-1} e^{-\imath k m} \varphi_\alpha^* (V_m-W) \langle m|\Psi,E\rangle\right]\right\}\\
& \quad =
\sum_{\alpha=1}^\wdth \int_{|z|=1} \frac{{\rm d}z}{2\pi \imath}\left\{\sum_{m=0}^{\lgth-1}
\frac{\varphi_\alpha\,z^{n-m}}{z^2+z(E+\imath \epsilon-\lambda_\alpha)+1}\,
\varphi_\alpha^* (V_m-W) \langle m|\Psi,E\rangle\right\}\\
& \quad =
\sum_{\alpha=1}^\wdth \sum_{m=0}^{\lgth-1} 
\frac{(\xi^\epsilon_\alpha)^{|n-m|}}
{\xi^\epsilon_\alpha-(\xi^\epsilon_\alpha)^{-1}}
\,\varphi_\alpha\varphi_\alpha^* (V_m-W) \langle m|\Psi,E\rangle,
\end{align*}
where $\xi^\epsilon_\alpha$ is the solution of
$z^2+z(E+\imath \epsilon-\lambda_\alpha)+1=0$ which is inside the unit disc.
Let $\alpha>s$, then the solutions for $\epsilon=0$ are $u_\alpha e^{\pm\gamma_\alpha}$
and $\xi^\epsilon_\alpha$ converges to $u_\alpha e^{-\gamma_\alpha}$ for
$\epsilon\to 0$.
For the elliptic channels $\alpha\leq s$ the solutions
for $\epsilon=0$ are $e^{\pm \imath k_\alpha}$ being both on the unit circle.
As $|\xi^\epsilon_\alpha|<1$ the sign of its imaginary part is different
to the sign of the imaginary part of
$\xi^\epsilon_\alpha+(\xi^\epsilon_\alpha)^{-1}=\lambda_\alpha-E-\imath \epsilon$.
Since $k_\alpha\in(0,\pi)$ this leads to
$\lim_{\epsilon\searrow 0}\xi^\epsilon_\alpha=e^{\imath k_\alpha}$
and $\lim_{\epsilon\nearrow 0}\xi^\epsilon_\alpha=e^{-\imath k_\alpha}$.
Hence by the calculations above and 
\eqref{eq-LipSch1}, \eqref{eq-LipSch2} we get
\begin{align}
\langle n|\Psi,E\rangle
=\;&
\langle n|\Psi_{\text{in}}^0,E\rangle
\,+\,
\sum_{\alpha\leq s}
\sum_{m=0}^{\lgth-1}
\frac{\varphi_\alpha e^{\imath k_\alpha |n-m|}}
{2\imath\sin(k_\alpha)}
\varphi_\alpha^* (V_m-W) \langle m|\Psi,E\rangle \notag \\
&+\,\sum_{\alpha>s}\sum_{m=0}^{\lgth-1}
\frac{\varphi_\alpha u_\alpha^n e^{-\gamma_\alpha|n-m|}}
{e^{-\gamma_\alpha}-e^{\gamma_\alpha}}
\varphi_\alpha^* (V_m-W) \langle m|\Psi,E\rangle
\end{align}
\begin{align}
\langle n|\Psi,E\rangle
=\;&
\langle n|\Psi_{\text{out}}^0,E\rangle
\,+\,
\sum_{\alpha\leq s}
\sum_{m=0}^{\lgth-1}
\frac{\varphi_\alpha e^{-\imath k_\alpha |n-m|}}
{2\imath\sin(k_\alpha)}
\varphi_\alpha^* (V_m-W) \langle m|\Psi,E\rangle \notag \\
&+\,\sum_{\alpha>s}\sum_{m=0}^{\lgth-1}
\frac{\varphi_\alpha u_\alpha^n e^{-\gamma_\alpha|n-m|}}
{e^{-\gamma_\alpha}-e^{\gamma_\alpha}}
\varphi_\alpha^* (V_m-W) \langle m|\Psi,E\rangle
\end{align}
Thus, we see that in the hyperbolic channels, the extended state $|\Psi,E\rangle$ has only exponential decaying parts which justifies
the definition for scattering states.
Moreover, if $|\Psi,E\rangle$ is the scattering state
associated to the coefficients
$a^+,a^-, b^+,b^-$ and $\hat{a}^+,\hat{b}^-$ as in 
\eqref{eq-scst1} and \eqref{eq-scst2}, then for $n\to\pm\infty$ the equations give
\begin{equation}\label{eq-Psi-in}
|\Psi^0_{\text{in}},E\rangle
\,=\,
\sum_{\alpha\leq s}\left[ a^+_\alpha |\Psi^0_\alpha,E,+\rangle
\,+\, e^{\imath k_\alpha \lgth} b^-_\alpha |\Psi^0_\alpha,E,-\rangle\,\right]
\end{equation}
and
\begin{equation}\label{eq-Psi-out}
|\Psi^0_{\text{out}},E\rangle
\,=\,
\sum_{\alpha\leq s} a^-_\alpha |\Psi^0_\alpha,E,-\rangle
\,+\, e^{-\imath k_\alpha \lgth} b^+_\alpha |\Psi^0_\alpha,E,+\rangle\;.
\end{equation}
Therefore, the scattering operator reduced to the energy shell can be
described by the $2s \times 2s$ matrix $\Ss^E$ 
defined by
\begin{equation}\label{eq-defS}
\Ss^E\,\left( \begin{matrix} a^+ \\ b^-
\end{matrix} \right)
\,=\,
\left( \begin{matrix}
a^- \\ b^+
\end{matrix}\right)\;.
\end{equation}
In particular, from \eqref{eq-redtr2} we obtain that $\widetilde\Tt^E=\Cc_s\hat\Tt^E \Cc_s^*\,\in\,{\rm U}(s,s)$ 
represents the $S$-transfer matrix.
The existence of $\Ss^E$ follows from \eqref{eq-redtr2} and the following theorem which is proved in Appendix~\ref{sec-scattering-matrix}.
\begin{theo} \label{th-T-S}
For any matrix $\widetilde\Tt\in{\rm U}(s,s)$ there is a unique unitary
matrix $S\in{\rm U}(2s)$ with the property that for any
$a^+,a^-, b^+,b^-\in\,\CC^s$ one has
\begin{equation}
\label{eq-relTS}
\widetilde{\Tt} \binom{a^+}{a^-}
\,=\,
\binom{b^+}{b^-}\quad\Leftrightarrow \quad
S \binom{a^+}{b^-}
\,=\,
\binom{a^-}{b^+}
\end{equation}
\end{theo}

\vspace{.2cm}

Putting equations \eqref{eq-in-out}, \eqref{eq-Psi-in}, \eqref{eq-Psi-out} and \eqref{eq-defS} 
together one can write the operator
$\Ss$ as an integral over the energy $E$.
The number of elliptic channels is a step function $s(E)$.
So far we considered one fixed energy and set the elliptic channels to be the ones for $\alpha=1,\ldots,s$.
But when varying $E$ one should take into account that the channels which are elliptic are different ones for
different energy intervals.
Therefore let $\alpha(E,1),\ldots,\alpha(E,s(E))$ denote the elliptic channels for $E$.
Correspondingly for pseudo-eigenstates satisfying $\Ss|\Psi^0_{\text{in}},E\rangle =|\Psi^0_{\text{out}},E\rangle$, 
define the coefficients $a^\pm_{\alpha(E,i)}$ and $b^\pm_{\alpha(E,i)}$.
Then the scattering matrix $\Ss^E$ satisfies \eqref{eq-defS} with 
$a^\pm =(a^\pm_{\alpha(E,1)},\ldots,a^\pm_{\alpha(E,s(E))})^\top$ and the analogue definitions for $b^\pm$.
Furthermore, let $e_{E,i,+}$ be the $i$-th and $e_{E,i,-}$ be the $(s(E)+i)$-th canonical basis vector of
$\CC^{2s(E)}$ for $i=1,\ldots,s(E)$. 
By \eqref{eq-Psi-in}, \eqref{eq-Psi-out} and \eqref{eq-defS}
the matrix element $e_{E,j,+}^* \Ss^E e_{E,i,+}$ corresponds to
the contribution of $a^+_{\alpha(E,i)}$ to $a^-_{\alpha(E,j)}$.
The meaning of the other matrix elements can also be read off these equations and one finally obtains the following.
\begin{proposi}
The scattering operator $\Ss$ is given by
\begin{equation}\label{eq-int-S}
\Ss=
\!\!\int \!\!dE
\left[ \sum_{\substack{{i,j=1,\ldots, s(E)} \\ {\ominus,\oslash\in\{+,-\}} }} \!\!\!\!\!\!
e^{\imath \theta(E,i,j,\ominus,\oslash)}| \Psi^0_{\alpha(E,j)},E,-\oslash \rangle
\left( e_{E,j,\oslash}^* \,\Ss^E\, e_{E,i,\ominus} \right)
\langle \Psi^0_{\alpha(E,i)},E,\ominus|
\right]
\end{equation}
where the correction phase $\theta(E,i,j,\ominus,\oslash)$ is given by
$$
\theta(E,i,j,\ominus,\oslash)\,=\,
-\delta_{\ominus,-}\,k_{E,i} \lgth \;-\;
\delta_{\oslash,-}\,k_{E,j} \lgth \;.
$$
with
$$
E=-2\cos(k_{E,i})+\lambda_{\alpha(E,i)}\;,\quad k_{E,i}\,\in\,(0,\pi)\;.
$$
\end{proposi}

\noindent The phase $\theta(E,i,j,\ominus,\oslash)$ comes from terms of the form $e^{\imath k_\alpha \lgth}$ 
appearing as factors in \eqref{eq-Psi-in} and \eqref{eq-Psi-out}. Now we can finally prove Theorem~\ref{th-main0}.

\vspace{.2cm}

\noindent {\bf Proof of Theorem~\ref{th-main0}.}
The direct integral $\int^\oplus \CC^{2s(E)}\,{\rm d}E$ is represented by functions $f(E)$ with $f(E)\in\CC^{2s(E)}$ and the scalar 
product is given by $\langle f | g \rangle = \int f(E)^* g(E){\rm d}E$.
Let us define the unitary operator $\Uu: \ell^2(\ZZ,\CC^\wdth) \to \int^\oplus \CC^{2s(E)} {\rm d}E$ by
\begin{equation}\label{eq-def-Uu}
(\Uu |n,j\rangle)(E) \,=\,
\sum_{i=1}^{s(E)} \sum_{\ominus\in\{+,-\}} e_{E,i,\ominus}\,
\langle \Psi^0_{\alpha(E,i)},E,\ominus|n,j\rangle
\end{equation}
and the diagonal $s(E)\times s(E)$ matrix $k_E$ by
\begin{equation}
k_E\,=\,\diag(k_{E,1},\ldots,k_{E,s(E)})\;.
\end{equation}
Using  \eqref{eq-delta} and \eqref{eq-part1E} one obtains that $\Uu$ is unitary.
The equations \eqref{eq-int-S} and \eqref{eq-sp-H_0} can be written as
\begin{align}
\Ss &\,=\, \Uu^*\, \left[\int^\oplus 
\begin{pmatrix} \nul & e^{-k_E\lgth} \\ \one & \nul \end{pmatrix}
\,\Ss^E\,\begin{pmatrix} \one & \nul \\ \nul & e^{-ik_E\lgth} \end{pmatrix}
{\rm d}E\right]\, \Uu\;, \label{eq-Ss-dint}\\
H_0&\,=\,
\Uu^* \, \left[\int^\oplus E\,\one_{2s(E)} {\rm d}E
\right]\,\Uu\;. \label{eq-H_0-sp}
\end{align}
A special case is $H=H_0$, where $\Ss=1$ and $\Ss^E=\left(\begin{smallmatrix}
                                                                 \nul & e^{ik_E\lgth} \\ e^{ik_E\lgth} & \nul
                                                                \end{smallmatrix} \right)$ and hence
\eqref{eq-Ss-dint} gives $\Uu^* \Uu = \one$.
Equations \eqref{eq-def-Uu}~-~\eqref{eq-H_0-sp} show Theorem~\ref{th-main0} part (i).
Part (ii) and (iii) follow from the equations
\eqref{eq-Tr-rel}, \eqref{eq-condred}, \eqref{eq-redtr0}, \eqref{eq-redtr2} and \eqref{eq-defS}. \hfill $\Box$

\section{$\Ss^E$ as limit of higher dimensional scattering matrices}
\label{sec-ideal}

In this section we prove Theorem~\ref{th-main}.
Recall that in the introduction we constructed an ideal lead described by the operator $H_I$ as in \eqref{eq-HI}.
The corresponding Hermitian matrix $W_I\in{\rm Her}(\wdth)$ was
defined by \eqref{eq-def-WI} which is equivalent to
\begin{equation} \label{eq-Delta_I}
 W_I \,=\, U{\rm diag}(\lambda_1,\ldots,\lambda_s,E,\ldots,E)\,U^*\,,
\end{equation}
where $U=(\varphi_1,\ldots,\varphi_\wdth)$ as in \eqref{eq-U}. 
The corresponding extended states of $H_I$ as well as the wave-numbers $k_\alpha$ for the energy $E$ are the same as the once of $H_0$ 
for $\alpha\leq s$ and given by \eqref{eq-E-k} and \eqref{eq-P0-E}. 
For $\alpha>s$ there are additional extended states of $H_I$ for the energy $E$ defined as in 
\eqref{eq-P0-E} with $k_\alpha=\frac{\pi}{2}$.

Inserting a piece of the cable $H_0$ of length $m$ followed by the scatterer and another piece of the cable of length $m$ into the ideal lead
$H_I$ is described by the operator $H(m)$ as defined in \eqref{eq-def-H(m)} (cf. Figure~\ref{fig1}).

Similar to above one can introduce the
vectors $a_I^\pm$ and $b_I^\pm$ in $\CC^\wdth$ describing a formal solution $(\Psi_n)_n$
of the eigenvalue equation $H(m)\Psi=E\Psi$ for $n<-m$ and $n\geq \lgth+m$, but this time there are only
elliptic channels.

The transfer matrix of the inserted piece given by the block of $H(m)$ from $n=-m$ to $n=L+m-1$ is given by
$\Tt(m)={(\Tt^{0,E})}^m \Tt^E_{0,\lgth} {(\Tt^{0,E})}^m$, with
$\Tt^{0,E}$ as defined in \eqref{eq-freeTr}.
To get the relation between $a_I^\pm$ and $b_I^\pm$ we have to follow the same steps as in Sections~\ref{sec-normal} and \ref{sec-reduced}.
Hence, let us introduce
the matrix $M_I$ similar to $M$ in \eqref{eq-defM} by
\begin{equation}
M_I=
\left( \begin{matrix}
U & \nul \\ \nul & U
\end{matrix}\right)
\left(
\begin{smallmatrix}
 (\sin(k))^{-1/2} & \nul & \nul & \nul \\
\nul & \one  & \nul & \nul \\
\cos(k)(\sin(k))^{-1/2} & \nul & (\sin(k))^{1/2} & \nul \\
\nul & \nul & \nul & \one 
\end{smallmatrix}
\right)\;.
\end{equation}
Then $M_I^{-1} \Tt(m) M_I \left(\begin{smallmatrix}a_I^+ + a_I^-\\ \imath(a_I^+-a_I^-)\end{smallmatrix}\right)=
\left(\begin{smallmatrix} b_I^+ + b_I^- \\ \imath (b_I^+-b_I^-) \end{smallmatrix} \right)$.
Let $a^\pm,\hat a^\pm, b^\pm, \hat b^\pm$ satisfy the relations as in \eqref{eq-Tr-rel}, i.e.
$$
M^{-1} \Tt^E_{0,\lgth} M\,
\left( \begin{smallmatrix}
a^+ + a^- \\
2\hat{a}^+ \\
\imath (a^- - a^+) \\ 
2\hat{a}^-
\end{smallmatrix}\right)
\,=\,
\left( \begin{smallmatrix}
b^+ + b^- \\
2\hat{b}^+ \\
\imath(b^- - b^+)\\ 
2\hat{b}^-
\end{smallmatrix}\right)\,.
$$
Then using \eqref{eq-freeTr-normal} one obtains
\begin{equation}\label{eq-T(m)-conj}
M^{-1} \Tt(m) M\,
\left( \begin{smallmatrix}
e^{-\imath mk} a^+ + e^{\imath mk}a^- \\
2\,u^m e^{-m\gamma} \,\hat{a}^+ \\
\imath (e^{\imath mk} a^- - e^{-\imath mk}a^+) \\ 
2\,u^m e^{m\gamma}\,\hat{a}^-
\end{smallmatrix}\right)
\,=\,
\left( \begin{smallmatrix}
e^{\imath mk}b^+ + e^{-\imath mk} b^- \\
2\,u^m e^{m\gamma}\,\hat{b}^+ \\
\imath(e^{-\imath mk}b^- - e^{\imath mk}b^+)\\ 
2 \,u^m e^{-m\gamma } \,\hat{b}^-
\end{smallmatrix}\right)\;.
\end{equation}
Furthermore by \eqref{eq-defM} and the definition of $M_I$ one has
\begin{equation}\label{eq-MI-M}
M_I^{-1} M 
\,=\,
\left(\begin{smallmatrix}
\one & \nul & \nul & \nul \\ 
\nul & u(2\sinh \gamma)^{-\frac12} &  \nul &  (2\sinh \gamma)^{-\frac12} \\
\nul & \nul & \one & \nul  \\ 
\nul & e^{-\gamma} (2\sinh \gamma)^{-\frac12} & \nul & u e^{\gamma} (2\sinh \gamma)^{-\frac12}
\end{smallmatrix}\right)\;.
\end{equation}
To simplify notations let us define 
\begin{equation}\label{eq-tl-a}
\begin{pmatrix}
   \tilde{a}_m^+ \\ \tilde{a}_m^-
  \end{pmatrix}
\,=\, \frac{1}{\sqrt{2\sinh\gamma}}
\begin{pmatrix}
 u & \one \\ e^{-\gamma} & ue^\gamma
\end{pmatrix}
\begin{pmatrix}
 e^{-m \gamma} \hat a^+ \\
e^{m\gamma} \hat a^-
\end{pmatrix}
\end{equation}
and 
\begin{equation}\label{eq-tl-b}
\begin{pmatrix}
   \tilde{b}_m^+ \\ \tilde{b}_m^-
  \end{pmatrix}
\,=\, \frac{1}{\sqrt{2\sinh\gamma}}
\begin{pmatrix}
 u & \one \\ e^{-\gamma} & ue^\gamma
\end{pmatrix}
\begin{pmatrix}
 e^{m \gamma} \hat b^+ \\
e^{-m\gamma} \hat b^-
\end{pmatrix}\;.
\end{equation}
Then the equations \eqref{eq-T(m)-conj}, \eqref{eq-MI-M}, \eqref{eq-tl-a} and \eqref{eq-tl-b} yield
\begin{equation}\label{eq-St-ideal}
\Cc_\wdth\,M_I^{-1} \Tt(m) M_I\,\Cc_\wdth^*
\left( \begin{smallmatrix}
e^{-\imath mk} a^+ \\
u^m (\tilde{a}_m^+ +\imath \tilde{a}_m^-) \\
e^{\imath mk} a^- \\ 
u^m(\tilde{a}_m^+ - \imath\tilde{a}_m^-)
\end{smallmatrix}\right)
\,=\,
\left( \begin{smallmatrix}
e^{\imath mk}b^+ \\
u^m(\tilde{b}_m^+ +\imath\tilde{b}_m^-) \\
e^{-\imath mk}b^- \\ 
u^m(\tilde{b}_m^- -\imath\tilde{b}_m^-)
\end{smallmatrix}\right)\;.
\end{equation}
By \eqref{eq-S-tr-ideal} the matrix on the left hand side of \eqref{eq-St-ideal} is equal to 
the $S$-transfer matrix describing the scattering of
$H(m)$ with respect to $H_I$. Let $\Ss^E_I(m)$ be the related scattering matrix and
let us also introduce a phase normalization and consider the matrices 
$$
\hat\Ss^E_I(m)
\,=\,
D^m \,\Ss^E_I(m)\,D^m\,,\quad\text{where}\quad
D\,=\,
\left(\begin{smallmatrix}
e^{-\imath k} & \nul & \nul & \nul \\
\nul & u & \nul & \nul \\
\nul & \nul & e^{-\imath k} & \nul \\
\nul & \nul & \nul & u \end{smallmatrix}\right)\;.
$$
Then \eqref{eq-St-ideal} and the relation between scattering and $S$-transfer matrix \eqref{eq-Ss-Tt} yield
\begin{equation}
\label{eq-Smatrices}
\hat \Ss^E_I(m)\,
\left( \begin{smallmatrix}
a^+ \\
\tilde a_m^+ + \imath \tilde a_m^-  \\
b^- \\
\tilde b_m^+ - \imath \tilde b_m^-  
\end{smallmatrix}\right)
\,=\,
\left( \begin{smallmatrix}
a^- \\
\tilde a_m^+ - \imath \tilde a_m^-  \\
b^+ \\
\tilde b_m^+ + \imath \tilde b_m^- 
\end{smallmatrix}\right)\;.
\end{equation}

As the unitary group is compact, there is at least one limit point of this
sequence, let us call such a limit point $\hat\Ss^E_I$. 
If $\gamma$ is a multiple of the unit matrix, i.e. $\gamma=\tilde\gamma\,\one$, then
multiplying \eqref{eq-Smatrices} by
$(2\sinh \tilde\gamma)^{\frac12}e^{-\tilde \gamma m}$ and taking the limit $m\to\infty$ along a sequence where
$\Ss^E_I(m)$ converges to $\hat\Ss^E_I(m)$ we obtain from
\eqref{eq-tl-a} and \eqref{eq-tl-b} that
\begin{equation}
\label{eq-hypblocks}
\hat{\Ss}^E_I\,\left(
\begin{smallmatrix} 0 \\ (1+\imath u e^\gamma) \hat a^- \\ 0 \\ (u - \imath e^{-\gamma}) \hat b^+
\end{smallmatrix}\right)
\,=\,
\left(\begin{smallmatrix}
0 \\ (1 -\imath u e^\gamma) \hat a^- \\ 0 \\ (u + \imath e^{-\gamma}) \hat b^+
\end{smallmatrix}\right)\;.
\end{equation}
As the reduced transfer matrix is supposed to exist, we 
can always choose
$\hat a^+$ to get any $\hat b^+$ we want and hence, 
$\hat a^-$ and $\hat b^+$ can be chosen independently.

If $\gamma$ is not a multiple of the unit matrix, then for each $\gamma_\alpha$ we 
take the entries of $\hat a^-$ and $\hat b^+$ to be zero which belong to a $\gamma_\beta$ greater than $\gamma_\alpha$.
Then we multiply \eqref{eq-Smatrices} by
$(2\sinh \gamma_\alpha)^{\frac12}e^{-\tilde \gamma_\alpha m}$ and take the limit $m\to\infty$.
Doing this for any $\gamma_\alpha$ we also obtain \eqref{eq-hypblocks} for any vectors $\hat a^-,\, \hat b^+$ by linearity.

Furthermore one can choose $\hat a^- = 0$ and tune $\hat a^+$ such
that $\hat b^+ = 0$. Then $\tilde a_m^-$ and $\tilde b_m^+$ converge both to $0$ and 
$a^\pm, b^\pm$  are related by the scattering matrix $\Ss^E$. Hence, in this case the limit $m\to \infty$ of \eqref{eq-Smatrices}
along an appropriate subsequence yields
\begin{equation}
\label{eq-ellblocks}
\hat\Ss^E_I \,\left(\begin{smallmatrix}
a^+ \\ 0 \\ b^- \\ 0
\end{smallmatrix}\right)
\,=\,
\left(\begin{smallmatrix}
a^- \\ 0 \\ b^+ \\ 0
\end{smallmatrix}\right)
\quad \Leftrightarrow \quad
\Ss^E \binom{a^+}{b^-}
\,=\,
\binom{a^-}{b^+}\;.
\end{equation}
These two equations, \eqref{eq-hypblocks} 
and \eqref{eq-ellblocks}, determine any limit point of $\hat \Ss^E_I(m)$ uniquely.
Therefore, the limit $\hat\Ss^E_I=\lim_{m\to \infty} \hat\Ss^E_I(m)$ exists and
there is a relation between $\hat\Ss^E_I$ and $\Ss^E$ given by
\begin{equation}
\Ss^E\,=\,
\left(\begin{matrix} R & T' \\ T & R' \end{matrix}\right)\;
\qquad \Leftrightarrow \qquad
\hat\Ss^E_I\,=\, 
\left(\begin{smallmatrix}
R & \nul & T' & \nul \\
\nul & -e^{i\theta} & \nul & \nul \\
T' & \nul & R' & \nul \\
\nul & \nul & \nul & e^{i\theta}
\end{smallmatrix}\right)\;.
\end{equation}
where 
\begin{equation}\label{eq-def-theta}
e^{i\theta}=\left[\sinh(\gamma)+\imath u\right][\cosh(\gamma)]^{-1}\,\in\,{\rm U}(\wdth-s)\,.
\end{equation}
This shows Theorem~\ref{th-main}. \hfill $\Box$

%


\appendix

\section{Proof of Theorem~\ref{th-T-S}}
\label{sec-scattering-matrix}

In this appendix we prove Theorem~\ref{th-T-S}. It states that for any matrix $\tilde\Tt\in{\rm U}(s,s)$ there is a unitary matrix 
$S\in{\rm U}(2s)$ such that
\begin{equation}
\label{eq-relTS2}
\widetilde{\Tt} \left(\begin{matrix} a^+ \\ a^- \end{matrix} \right)
\,=\,
\left(\begin{matrix} b^+ \\ b^- \end{matrix} \right) \quad\Leftrightarrow \quad
S \left(\begin{matrix} a^+ \\ b^- \end{matrix} \right)
\,=\,
\left(\begin{matrix} a^- \\ b^+ \end{matrix} \right)\;.
\end{equation}

Let $\widetilde{\Tt}=\left(\begin{smallmatrix} A & B \\ C & D \end{smallmatrix}\right)$.
We first prove the existence and then uniqueness of $S$.
As $\widetilde{\Tt}\in{\rm U}(s,s)$ one has
\begin{equation}
\label{eq-str}
\begin{matrix}
A^*A &=& \one+C^*C&, \\
D^*D &=& \one+ B^*B&, \\
A^*B &=& C^*D\;\;\;\;&,
\end{matrix}
\qquad\qquad
\begin{matrix}
AA^*&=&\one+BB^* \\
DD^*&=&\one+CC^*\\
AC^*&=&BD^*\;\;\;\;.
\end{matrix}
\end{equation}
As $A^*A\geq \one$ there exists a unitary matrix $U_{l,+}$ and a 
real diagonal matrix $Q\geq \one$ such that $A^*A=U_{l,+}^*QU_{l,+}$.
Define $U_{r,+}$ by
\begin{equation}
\label{eq-A}
U_{r,+}
\,=\,
AU_{l,+}^* \sqrt{Q^{-1}}\qquad \Leftrightarrow\qquad
A\,=\,U_{r,+}\sqrt{Q} U_{l,+}
\end{equation}
Then $U_{r,+}^*U_{r,+}=\sqrt{Q^{-1}}U_{l,+}A^*AU_{l,+}^*\sqrt{Q^{-1}}=\one$ and hence $U_{r,+}$ is unitary.
Furthermore one has $C^*C=A^*A-\one = U_{l,+}^*(Q-\one)U_{l,+}$.
Hence there exists $U_{r,-}\in {\rm U}(s)$ such that
\begin{equation}
\label{eq-C}
U_{r,-} \sqrt{Q-\one}
\,=\,CU_{l,+}^*\qquad\Leftrightarrow\qquad
C\,=\,U_{r,-} \sqrt{Q-\one} U_{l,+}
\end{equation}
$U_{r,-}$ is uniquely determined if $Q-\one$ is
invertible, otherwise it is not.
Now define $U_{l,-}$ by
\begin{equation}
\label{eq-D}
 U_{l,+}=
\sqrt{Q^{-1}} U_{r,-}^* D
\qquad\Leftrightarrow\qquad
D\,=\,
 U_{r,-}\sqrt{Q}U_{l,-}
\end{equation}
Then using \eqref{eq-str} and \eqref{eq-C} one finds
$
 U_{l,-} U_{l,-}^*
=\one
$
and hence $U_{l,-}$ is also unitary.
Furthermore one obtains
using \eqref{eq-str}, \eqref{eq-A}, \eqref{eq-C} and \eqref{eq-D}
\begin{align}
B
&\,=\,
(A^*)^{-1} C^*D
\,=\, (U_{l,+}^*\sqrt{Q}U_{r,+}^*)^{-1} U_{l,+}^*\sqrt{Q-\one}U_{r,-}^* U_{r,-}
\sqrt{Q} U_{l,-}\nonumber \\
&\,=\,
U_{r,+}\sqrt{Q^{-1}}\sqrt{Q-\one}\sqrt{Q} U_{l,-}
\,=\,U_{r,+} \sqrt{Q-\one}{U_{l,-}}
\label{eq-B}
\end{align}
Now using \eqref{eq-A}, \eqref{eq-C}, \eqref{eq-D} and \eqref{eq-B} one obtains
\begin{equation}
\label{eq-Tt}
\widetilde\Tt
\,=\,
\left(\begin{matrix}
U_{r,+} & \nul \\ \nul & U_{r,-} \end{matrix}\right)
\left(\begin{matrix}
\sqrt{Q} & \sqrt{Q-\one} \\ \sqrt{Q-\one} & \sqrt{Q}
\end{matrix}\right)
\left(\begin{matrix}
U_{l,+} & \nul \\ \nul & U_{l,-} \end{matrix}\right)
\end{equation}
As $Q\geq\one$, one has $\one-Q^{-1}\geq \nul$ and hence
$\sqrt{\one-Q^{-1}}$ is a well-defined, non-negative diagonal matrix.
Thus we can define the unitary Matrix
\begin{equation}
\label{eq-Ss}
S
\,=\,
\left(\begin{matrix}
U_{l-}^* & \nul \\ \nul & U_{r,+} \end{matrix}\right)
\left(\begin{matrix}
-\sqrt{\one-Q^{-1}} & \sqrt{Q^{-1}} \\ 
\sqrt{Q^{-1}} & \sqrt{\one-Q^{-1}} \end{matrix}\right)
\left(\begin{matrix}
U_{l,+} & \nul \\ \nul &  U_{r,-}^*
\end{matrix}\right)
\end{equation}
It is now easy to check, that $\Ss$ fulfills \eqref{eq-relTS2}.

\vspace{.3cm}

To prove uniqueness of $S$ assume $\hat S$ also
fulfills \eqref{eq-relTS2}. But then \eqref{eq-relTS2} implies for any vector $v\in\CC^{2s}$
that $Sv=\hat S v$ and hence $S=\hat S$.
\hfill $\Box$

\vspace{.5cm}

\noindent {\bf Remark.} {The converse is not true. One cannot find a matrix $\widetilde\Tt$ for all unitary matrices 
$\Ss\in{\rm U}(2s)$ such that the relation above is fulfilled. 
Looking at the block structure $\Ss=\left(\begin{smallmatrix}
                                           R & T' \\ T & R
                                          \end{smallmatrix}\right)$
The matrix $\widetilde{\Tt}$ exists if $T$ is invertible {\rm (}which is equivalent to $T'$ being invertible{\rm )}.
$T$ and $T'$ are related to the transfer of waves. If they are not invertible, then 
there is one planar wave which is totally reflected by the scatterer. Hence a transfer
does not occur for this wave and the transfer matrix is not defined.}

\end{document}